\documentclass[aps,prl,endfloat,preprint,superscriptaddress]{revtex4}

\usepackage{graphicx}

\newcommand{\omegaL}{\omega_\mathrm{L}}
\newcommand{\omegaLO}{\omega_\mathrm{LO}}
\newcommand{\omegaS}{\omega_\mathrm{S}}

\newcommand{\TE}{\tau_\mathrm{e}}
\newcommand{\ELO}{E_\mathrm{LO}}
\newcommand{\EO}{E_\mathrm{O}}
\newcommand{\EL}{E_\mathrm{L}}
\newcommand{\AL}{A_{\rm L}}

\newcommand{\AO}{A_{\rm O}}
\newcommand{\ALO}{A_{\rm LO}}

\newcommand{\DHA}{X}
\newcommand{\Hs}{X _{\rm s}}
\newcommand{\Ht}{X _{\rm t}}
\newcommand{\Hsf}{X _{\rm sf}}

\newcommand{\As}{A _{\rm s}}
\newcommand{\At}{A _{\rm t}}

\newcommand{\Ys}{Y _{\rm s}}
\newcommand{\Yt}{Y _{\rm t}}
\newcommand{\Ysf}{Y _{\rm sf}}
\newcommand{\Yf}{Y _{\rm f}}
\newcommand{\BWPM}{\Delta \omega _{\rm PM}}

\newcommand{\BWF}{\Delta \omega_\mathrm{F}}
\newcommand{\SONE}{S _{ 1}}
\newcommand{\STWO}{S _{ 2}}
\newcommand{\ttau}{(t+\tau)}

\newcommand{\sinc}{{\rm sinc}}
\newcommand{\RE}{{\rm Re}}

\begin{document}

%\preprint{}

\title{Spatiotemporal heterodyne detection}
\author{Michael Atlan}
\email[Corresponding author: ]{atlan@lkb.ens.fr}
\author{Michel Gross}
%\thanks{}
%\affiliation{}
\affiliation{Laboratoire Kastler Brossel, \'Ecole Normale
Sup\'erieure, Universit\'e Pierre et Marie-Curie - Paris 6, Centre
National de la Recherche Scientifique, UMR 8552; 24 rue Lhomond,
75005 Paris, France}

\date{\today}

\begin{abstract}

We describe a scheme into which a camera is turned into an efficient
tunable frequency filter of a few Hertz bandwidth in an off-axis,
heterodyne optical mixing configuration, enabling to perform
parallel, high-resolution coherent spectral imaging. This approach
is made possible through the combination of a spatial and temporal
modulation of the signal to reject noise contributions. Experimental
data obtained with dynamically scattered light by a suspension of
particles in brownian motion is interpreted.

\end{abstract}



\maketitle

\section{Introduction}

\subsection{Coherent spectral imaging}

Coherent spectroscopy enables one to study mechanisms involving
dynamic light scattering. The spectral distribution of a
monochromatic optical field scattered by moving particles is
modified as a consequence of momentum transfer (Doppler broadening).
The measurement of the Doppler linewidth of this field (referred as
object field) with an optimal sensitivity is crucial, since Doppler
conversion yields are typically low.\\

\begin{figure}[H]%[H]figs a la fin
  \begin{center}%
   \includegraphics[width = 8.0cm]{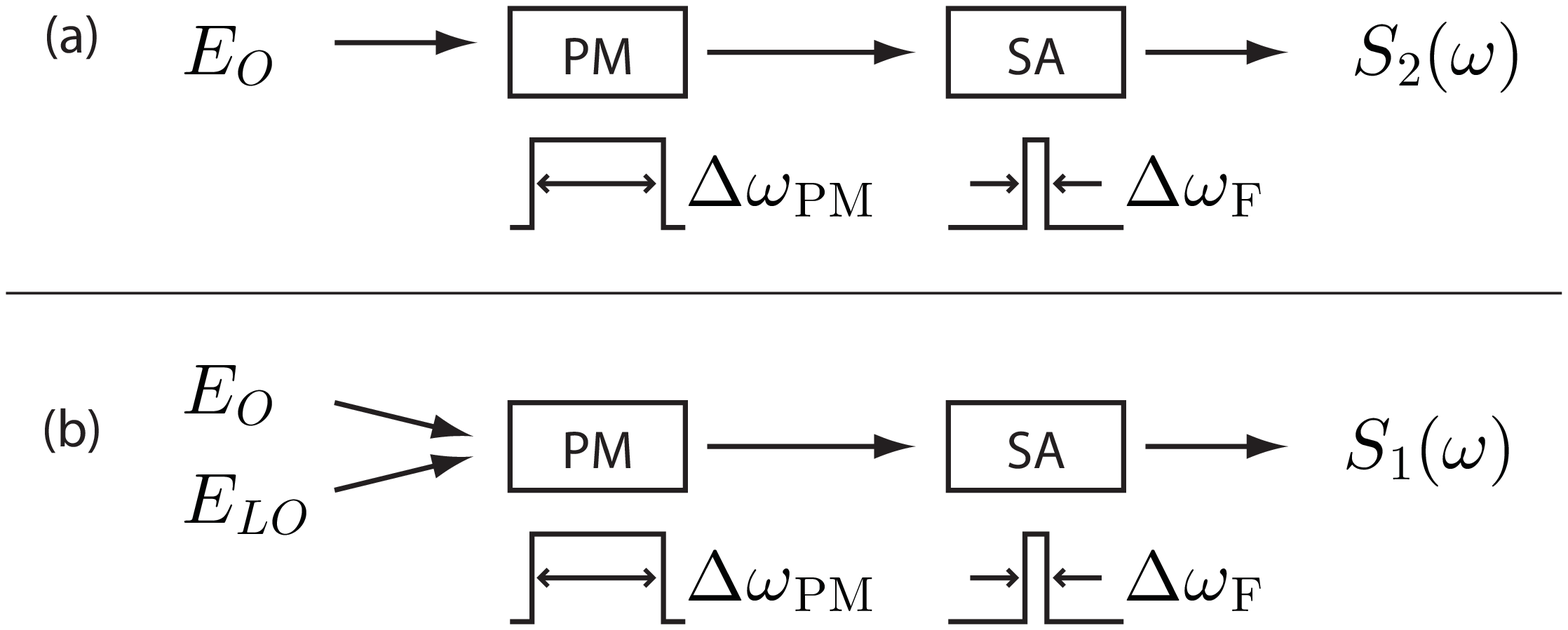}
\caption{Coherent spectral detection schemes : homodyne (a) and
heterodyne (b) optical mixing. PM : photomixer (square-law
detector). SA : spectrum analyzer. $\EO$ object field. $\ELO$ :
local oscillator field. $\SONE$, $\STWO$ : first and second order
object field spectral distributions.\label{fig_coh_detex}}
  \end{center}
\end{figure}

Optical mixing (or \emph{postdetection filtering}) techniques are
derived from RF spectroscopy techniques \cite{Forrester1961}. They
can be grouped in two categories \cite{bk_BernePecora2000,
Brown1983} : homodyne and heterodyne schemes. In homodyne mixing
(fig. \ref{fig_coh_detex}(a)), self-beating object light impinges on
a $\BWPM$-bandwidth photodetector (also referred as optical mixer or
photo-mixer, PM). To assess a frequency component of the object
light, the output of the PM is sent to a spectrum analyser, whose
bandwidth $\BWF$ defines the detection resolution. The resulting
spectra are proportional to the second-order object field spectral
distribution \cite{bk_BernePecora2000, Brown1983} $\STWO (\omega)$.
In heterodyne mixing, sketched in fig. \ref{fig_coh_detex}(b), the
object light is mixed onto a PM with a frequency-shifted reference
beam, also called local oscillator (LO). The LO field ($\ELO$) is
detuned to provoke a heterodyne beat of the object-LO field cross
contributions to the recorded intensity. This beat is sampled by a
PM and sent to a spectrum analyser whose bandwidth $\BWF$ defines
the apparatus resolution. This scheme enables one to measure the
first order object field spectral distribution $\SONE (\omega)$
\cite{bk_BernePecora2000, Brown1983}.\\

In heterodyne optical mixing experiments, the PM bandwidth $\BWPM$
defines the span of the measurable spectrum (usually $\sim 1$ GHz).
The resolution $\BWF$ can be lowered down to the sub Hertz range,
which is suitable for, among other applications, spectroscopy of
liquid and solid surfaces \cite{Chung1997}, dynamic light scattering
\cite{bk_BernePecora2000} and in vivo laser Doppler anemometry
\cite{Stern1977}. But these schemes are inadequate for imaging
applications, because measurements are done on one point. The
absence of spatial resolution has been sidestepped by scanning
techniques \cite{EssexByrne1991, Briers2001}, designed at the
expense of temporal resolution.\\

We present a heterodyne optical mixing detection scheme on an array
detector (configuration of fig. \ref{fig_coh_detex}(b)). Typical
array detectors sampling rates seldom run over 1 kHz, failing to
provide a bandwidth large enough for most Doppler applications to
date. Nevertheless, their strong advantage is to perform a parallel
detection over a large number of pixels. We present a spatial and
temporal modulation scheme (spatiotemporal heterodyning) that uses
the spatial sampling capabilities of an area detector to
counterbalance the noise issue of a measurement in heterodyne
configuration performed in the low temporal frequency range (e.g.
with a typical CCD camera). The issue of using narrow-bandwidth
camera PMs is alleviated by detuning the LO field optical frequency
accordingly to the desired spectral point of the object field to
measure. Post-detection filtering results from a
numerical Fourier transform over a limited number of acquired images.\\

This heterodyne optical mixing scheme on a low frame rate array
detector can be used as a filter to analyze coherent light. It has
already been used in several applications yet, including the
detection of ultrasound-modulated diffuse photons
\cite{GrossAl-Koussa2003, Atlan2005}, low-light spectrum analysis
\cite{GrossDunn2005}, laser Doppler imaging \cite{AtlanGross2006RSI,
AtlanGrossVitalis2006, AtlanGrossLeng2006}, and dynamic coherent
backscattering effect study \cite{LesaffreAtlan2006}. The purpose of
this paper is to present its mechanism.

\subsection{Time domain description of the fields}

We consider the spatially and temporally coherent light field of a
CW, single axial mode laser (dimensionless scalar representation) :
\begin{equation}\label{eq_EL}
    \EL (t) = \AL \At(t) \exp \left( i \omegaL t + i \phi (t) \right)
\end{equation}
where $\omegaL$ is the angular optical frequency, $\AL$ is the
amplitude of the field (positive constant), $\At(t) = 1+a(t)$;
$|a(t)| \ll 1$ describes the laser amplitude
fluctuations and $\phi (t)$ the phase fluctuations. This
field shines a collection of scatterers that re-emits the object field
(or scattered field), described by the following
function :
\begin{equation}\label{eq_EO}
    \EO (t) = \AO \Ht(t) \At(t) \exp \left( i \omegaL t + i \phi (t) \right)
\end{equation}
where $\AO$ is a positive constant and $\Ht(t)$ is the time-domain
phase and amplitude fluctuation induced by dynamic scattering of the
laser field, i.e. the cause of the object field fluctuations due to
dynamic scattering we intend to study. As in conventional heterodyne
detection schemes, a part of the laser field, taken-out from the
main beam constitutes the reference (LO) field :
\begin{equation}\label{eq_ELO}
    \ELO (t) = \ALO \At(t) \exp \left( i \omegaLO t + i \phi (t) \right)
\end{equation}
where $\ALO$ is a positive constant. The LO optical frequency is
shifted with respect to the main laser beam by $\omegaLO - \omegaL$
to provoke a tunable temporal modulation of the interference pattern
resulting from the mix of the scattered and reference fields. It can
be done experimentally by diffracting the reference beam
with RF-driven Bragg cells (acousto-optic modulators) for example.\\

The expression of the light instant intensity impinging on the
camera PMs is :
\begin{equation}\label{eq_i_t_defn}
     i(t) \simeq i_0 (\RE [ E ]) ^2
\end{equation}
where $i_0$ is a positive constant and $\RE [ E ]$ is the real part
of the relevant optical complex field $E$. The PMs are considered
point-like, their antenna properties \cite{Siegman1966} are out of
the scope of this paper. To take into account the frequency
filtering implication of finite time-domain integration, the average
intensity detected by the square-law camera PMs is :
\begin{equation}\label{eq_I_t_defn2}
    I(t) = I_0 \int _{-\TE /2} ^{\TE /2} (E \ttau + E ^* \ttau)^2  \, {\rm d} \tau
\end{equation}
where $I_0$ is a positive constant and $\TE$ the exposure time. In
the frequency-domain, this integration corresponds to a sinc-shaped
low-pass filter function whose bandwidth (defined as the distance
between the peak and the first zero) is $1/\TE$.

\section{Optical configuration}

\subsection{Setup}

The common name for optical mixing experiments on an array detector
is digital holography. The original underlying interference
technique was invented to improve electron microscopy resolution and
was successfully applied to monochromatic light imaging
\cite{Gabor1948}. Since the availability of lasers and then digital
cameras, digital holography \cite{Goodman1967} has become an
integrant part of many coherent imaging schemes, for its propensity
to record both quadratures of an optical field.\\

\begin{figure}[h]%[h]
  \begin{center}%
   \includegraphics[width = 8.0cm]{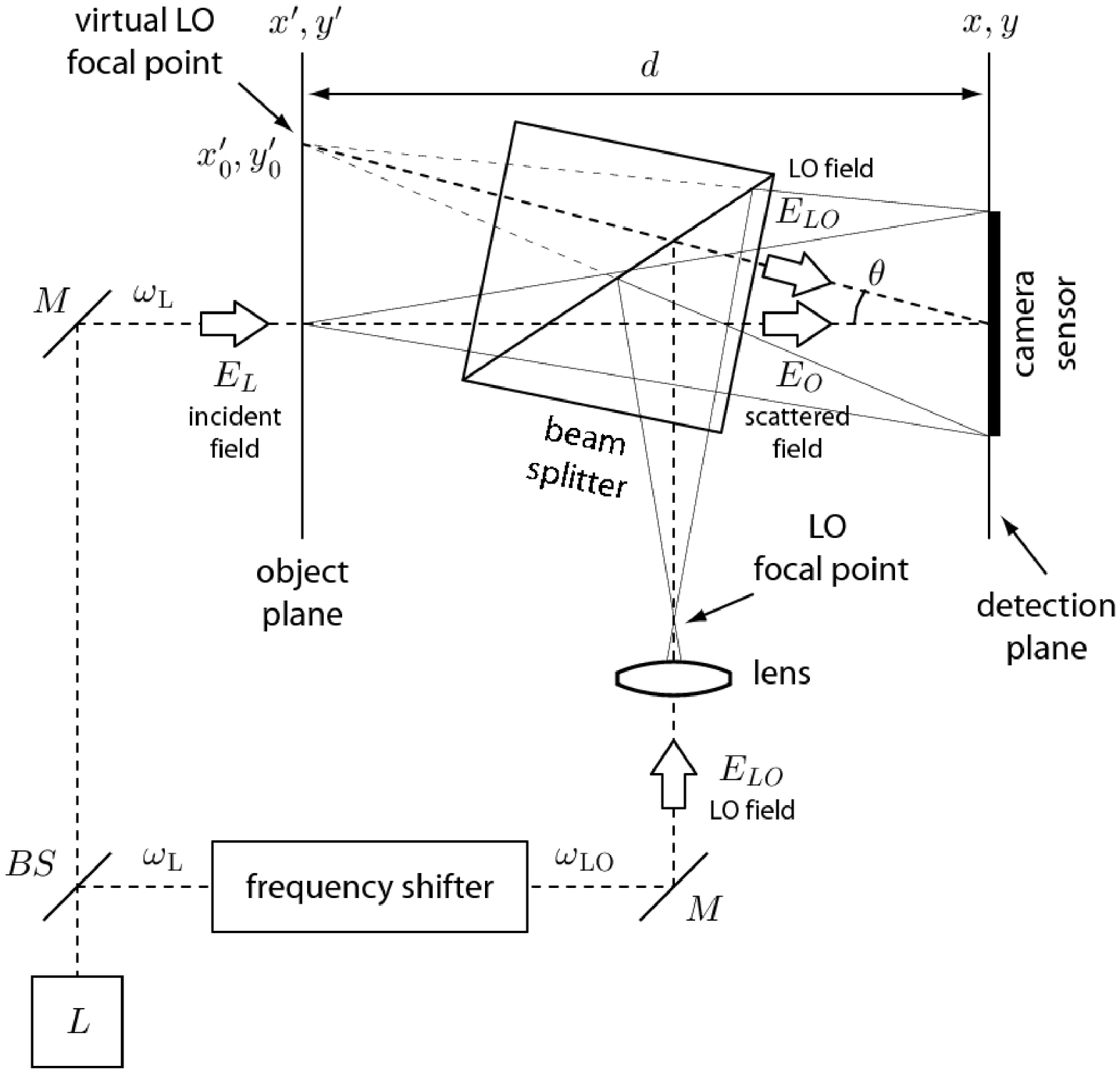}
\caption{Off-axis lensless Fourier configuration for heterodyne
holography. $L$ : laser. $M$ : mirror. $BS$ : beam splitter.
\label{fig_setup_config}}
  \end{center}
\end{figure}

The optical configuration we use can be either a lensless Fourier
off-axis holography setup \cite{Stroke1965, Wagner1999} or a Fresnel
off-axis holography setup \cite{Schnars1994}. In the lensless setup,
sketched in fig. \ref{fig_setup_config}, and considered throughout
this study, the point source of the spherical reference wave (LO
focal point) is located at $(x'_0,y'_0)$ in the plane of the object
to match the average curvature of the object field. An interference
pattern is recorded in the camera plane $(x,y)$. The reconstruction
algorithm used to calculate the spatial distribution of the
scattered field in the object plane consists of only one spatial
fast Fourier transform (FFT) \cite{Wagner1999}. This setup is used
to avoid the need of a general Fresnel reconstruction algorithm.  In
the Fresnel setup \cite{Schnars1994}, the LO is a plane wave and the
image reconstruction requires the use of numerical lenses and at
least two successive spatial domain FFTs \cite{Schnars2002}.

The digital hologram is recorded with the frequency-shifting method
introduced in the heterodyne holography technique
\cite{LeClercGross2000}, used in several imaging and detection
schemes \cite{GrossDunn2005, Atlan2005, AtlanGrossVitalis2006,
GrossAl-Koussa2003} for its ability to discriminate efficiently a
few Doppler-shifted photons, according to their frequency, from
background light \cite{GrossDunn2005, GrossAl-Koussa2003}. The
frequency-shifting method was chosen as an accurate
\cite{LeClercGross2000} dynamic phase-shifting detection scheme.
This method is derived from static phase-shifting interferometry
\cite{Creath1985} applied to phase-shifting holography
\cite{Yamaguchi1997}. It consists of detuning the LO optical
frequency with respect to the main laser beam, to provoke a time
domain modulation of the intensity pattern resulting from the cross
contribution of object and LO fields.

\subsection{Spatial and temporal dependencies}

Space (subscript s) and time (subscript t) dependencies of $\DHA$
and $A$ are noted $\DHA(x,y,t) = \Hs (x,y) \cdot \Ht (t)$, $A(x,y,t)
= \As (x,y) \cdot \At (t)$. Spatial frequencies (subscript sf) and
temporal frequencies (subscript f) reciprocal distributions form
Fourier pairs with time domain and spatial domain distributions. We
have :
\begin{equation}\label{eq_defn_TF_t}
\Yf(\omega) = \int _{-T/2} ^{T/2} \Yt(t) \exp(-i \omega t) \, {\rm
d}t
\end{equation}
and
\begin{eqnarray}\label{eq_defn_TF_xy}
\nonumber \Ysf(\xi,\eta) = \int _{-\Delta x/2} ^{\Delta x/2}\int
_{-\Delta y/2} ^{\Delta y/2} \Ys(x,y) \\
\times \exp(-2i\pi (x\xi+y\eta)) \, {\rm d}x \,{\rm d}y
\end{eqnarray}
where $Y$ is the considered distribution, $T$ is the measurement
time, $\Delta x$ and $\Delta y$ are the camera sensor widths.\\

In the lensless Fourier configuration, the LO wave curvature matches
the average curvature of the object field. Therefore, the heterodyne
interference pattern on the detector is formally \emph{equivalent}
\cite{Stroke1965} to the one that would impinge on the detector
under Fraunhofer conditions. Consequently, we can consider the (far
field diffraction) equivalent situation where the LO wave is a
tilted plane wave in the detector plane and the object field
distribution is the Fourier transform of its spatial distribution in
the object plane. The object and LO complex optical fields take the
following form:
\begin{eqnarray}\label{eq_EO_separated}
    \nonumber \EO(x,y,t)=\\
    \nonumber \AO \Hs(x,y)\As(x,y)\\
    \times \Ht(t)\At(t) \exp \left( i \omegaL t + i \phi (t) \right)
\end{eqnarray}
\begin{eqnarray}\label{eq_ELO_separated}
    \nonumber \ELO(x,y,t)=\\
    \nonumber \ALO \As(x,y) \exp \left(2i\pi (x \xi _0 + y \eta _0)
    \right)\\
    \times \At(t) \exp \left( i \omegaLO t + i \phi (t) \right)
\end{eqnarray}
where $\xi _0 = x'_0 / (\lambda d)$ and $\eta _0 = y'_0 / (\lambda
d)$. The tilt angle of the reference field leads to the phase factor
$\exp \left(2i\pi (x \xi _0 + y \eta _0) \right)$.
%$\exp(i\psi)$ is a quadratic phase factor defining the average curvature of the fields impinging on the detector.
The detected intensity is :
\begin{eqnarray}\label{eq_I_t_hetero1}
    \nonumber I(x,y,t) =\\
    I_0 \nonumber\int _{-\TE/2} ^{\TE/2} [ \EO (x,y,t+\tau) + \ELO
    (x,y,t+\tau)\\
    +\EO ^* (x,y,t+\tau) + \ELO ^*
    (x,y,t+\tau) ]^2 \, {\rm d} \tau
\end{eqnarray}
Expression \ref{eq_I_t_hetero1} has 16 terms among which 10 vanish
because of the presence of optical frequencies in the phase factors
\cite{Chung1997}, which lead to oscillations outside the detection
bandwidth. It can be rewritten as :
\begin{eqnarray}\label{eq_I_xyt_hetero0}
    \nonumber I(x,y,t) / I_0 =\\
     \nonumber \AO \ALO \, I_1(x,y) \cdot I_1(t)\\
    \nonumber + \AO \ALO \, I_2(x,y) \cdot I_2(t)\\
    \nonumber + \ALO^2 \, I_3(x,y) \cdot I_3(t)\\
    + \AO^2 \, I_4(x,y) \cdot I_4(t)
\end{eqnarray}
where the time-domain fluctuations of the intensity are :
\begin{eqnarray}\label{eq_I_t1defn}
    \nonumber I_1(t) = \int _{-\TE/2} ^{\TE/2} \Ht \ttau \left| \At \ttau \right|^2\\
     \times \exp \left( -i (\omegaLO - \omegaL)
     \ttau \right) \, {\rm d} \tau
\end{eqnarray}
\begin{eqnarray}\label{eq_I_t2defn}
    \nonumber I_2(t)  = \int _{-\TE/2} ^{\TE/2} \Ht ^* \ttau \left| \At \ttau \right|^2\\
     \times \exp \left( i (\omegaLO - \omegaL)
     \ttau \right)  \, {\rm d} \tau
\end{eqnarray}
\begin{eqnarray}\label{eq_I_t3defn}
I_3(t)  =  \int _{-\TE/2} ^{\TE/2} \left| \At \ttau \right|^2 \,
{\rm d} \tau
\end{eqnarray}
\begin{eqnarray}\label{eq_I_t4defn}
I_4(t)  = \int _{-\TE/2} ^{\TE/2} \left| \Ht \ttau \right|^2 \left|
\At \ttau \right|^2  \, {\rm d} \tau
\end{eqnarray}
and the spatial contributions to the intensity take the following
form :
\begin{eqnarray}\label{eq_I_xy_hetero1}
    \nonumber I_1 (x,y)  =\\
    \left|\As(x,y)\right|^2 \Hs(x,y) \exp \left( 2i\pi (x \xi _0 + y \eta _0) \right)
\end{eqnarray}
\begin{eqnarray}\label{eq_I_xy_hetero2}
    \nonumber I_2 (x,y)  =\\
    \left|\As(x,y)\right|^2 \Hs ^* (x,y) \exp \left( -2i\pi (x \xi _0 + y \eta _0) \right)
\end{eqnarray}
 where $\xi _0 = x'_0 / (\lambda d)$ and $ \eta _0
= y'_0 / (\lambda d)$.
\begin{eqnarray}\label{eq_I_xy_hetero3}
    I_3 (x,y) = \left|\As(x,y)\right|^2
\end{eqnarray}
\begin{eqnarray}\label{eq_I_xy_hetero4}
    I_4 (x,y) =  \left|\Hs (x,y)\right|^2 \left|\As(x,y)\right|^2
\end{eqnarray}
Temporal and spatial dependencies can be treated separately. Eq.
\ref{eq_I_xyt_hetero0} has four terms : the first two are object-LO
field cross terms. The third and fourth terms correspond to the LO
and object fields self beating contributions. The two first spatial
contributions, $I_1 (x,y)$ \& $I_2 (x,y)$ will be referred as
heterodyne, and $I_3 (x,y)$ \& $I_4 (x,y)$ as homodyne, as
well as their temporal counterparts.\\

An assumption about the relative amplitudes of the fields is usually
made \cite{bk_BernePecora2000} : the object field amplitude is much
lower than the LO field amplitude (heterodyne regime) :
\begin{equation}\label{eq_regime_hetero}
    \AO \ll \ALO
\end{equation}
This assumption leads to neglecting the object field self-beating
contribution, because its relative amplitude with respect to the
cross terms is $\AO / \ALO \ll 1$. Additionally, in usual optical
mixing schemes on photodiodes, laser amplitude fluctuations are
neglected (i.e. $a (t) = 0$) \cite{bk_BernePecora2000, Chung1997},
since the measurement is done with detectors whose bandwidth is
large enough to make this assumption valid, as the detection is made
at higher frequencies. But with a slow PM such as a CCD camera, the
frequency-domain noise resulting from these fluctuations is not
negligible.\\

A sequence of $n$ images is sampled and leads to a data cube
$I(x,y,t)$, acquired for a given detuning frequency
$\omegaLO-\omegaL$. To make a spatial map of one frequency component
of the object field, a demodulation in the time and spatial domains
is performed. As we will see, the heterodyne holography scheme
allows encoding of the spectral and spatial information about the
object field in the reciprocal space of the data cube, and it also
allows spatial discrimination of the self beating terms from the
object-LO field cross terms.

\section{Signal demodulation in the time domain.}

$I_1$ and $I_2$ are referred as the heterodyne terms (modulated at
the detuning frequency $\omegaLO - \omegaL$, on purpose). They
correspond to the object and LO field cross terms. $I_3$ and $I_4$
are referred as the homodyne terms (not \emph{explicitly} modulated
at the detuning frequency). These terms correspond to the object
field (and the LO field) self beating intensity contributions.\\

To measure a spectral component of the object field, $n$ samples of
$I(x,y,t)$ along the time axis are acquired. These images are used
to calculate both quadratures of the object field by making a time
domain demodulation consisting of calculating the first harmonic
component ($I(x,y,\omegaS / n)$) of the recorded sequence. This
method is also used in temporal heterodyne inline holography
\cite{Indebetouw1999, Indebetouw2001}. The relation between the
result of this demodulation and the temporal frequency content of
the object field is investigated.

\subsection{First heterodyne term}
Let's consider the first heterodyne term :
\begin{eqnarray}\label{eq_I1}
    I_1(t) =   F(t) \exp (-i (\omegaLO - \omegaL) t )
\end{eqnarray}
where $F(t)$ is :
\begin{eqnarray}\label{eq_F_t}
    \nonumber F(t) = \int _{-\TE/2} ^{\TE/2} \left| \At \ttau \right|^2 \Ht \ttau  \\
    \times \exp (-i (\omegaLO - \omegaL) \tau ) \, {\rm d} \tau
\end{eqnarray}
According to eq. \ref{eq_F_t}, $F(t)$ corresponds to the result of
the application of a bandpass filter (of center
frequency $\omegaLO - \omegaL$ and bandwidth $1/\TE$) on $\Ht(t)$.\\

To \emph{demodulate} the signal, the first harmonic ($\omegaS/n$) of
the sampled image sequence is calculated by FFT :
\begin{eqnarray}\label{eq_Itilde1}
I_1(\omegaS/n) = \sum  _{k=1} ^{n} I_1(t_k) \, \exp (-2 i k \pi /
n )
\end{eqnarray}
where $t_k = 2 k \pi / \omegaS$ is the instant at which the $k^{\rm
th}$ image is recorded. We have :
\begin{eqnarray}\label{eq_Itilde1_bis}
\nonumber I_1(\omegaS/n) = \sum  _{k=1} ^{n} F(t_k) \, \exp (-i
(\omegaLO - \omegaL) t_k ) \\ \times \exp (-2 i k \pi / n )
\end{eqnarray}
$I_1(\omegaS/n)$ is a measurement of the $\omegaLO - \omegaL +
\omegaS/ n$ frequency component of $F(t)$ (the corresponding
instrumental bandwidth is $\omegaS/n$). Eq. \ref{eq_F_t} and
\ref{eq_Itilde1_bis} show that the spectral resolution of this
measurement should either be limited by the measurement time
($n/\omegaS$) and/or the spectral distribution of the amplitude
noise of the laser.

\subsection{Second heterodyne term}

The assessment of the quantity $I_2(\omegaS/n)$ is straightforward
:
\begin{eqnarray}\label{eq_Itilde2_bis}
\nonumber I_2(\omegaS/n) = \sum  _{k=1} ^{n} F^{*}(t_k) \, \exp (i
(\omegaLO - \omegaL) t_k ) \\ \times \exp (-2 i k \pi / n )
\end{eqnarray}
where $F^*(t)$ is the complex conjugate of $F(t)$ defined by eq.
\ref{eq_F_t}. Equations \ref{eq_F_t} and \ref{eq_Itilde2_bis}
define the action of two successive selective filters on $\Ht(t)$.
The filters bandwidths are the same as the ones introduced to
describe the first heterodyne term. Their action is presented and
compared to their counterparts appearing in the first homodyne
term in the next section.

\subsection{Bandpass filtering}

\begin{figure}[h]%[h]
  \begin{center}%
   \includegraphics[width = 8.0cm]{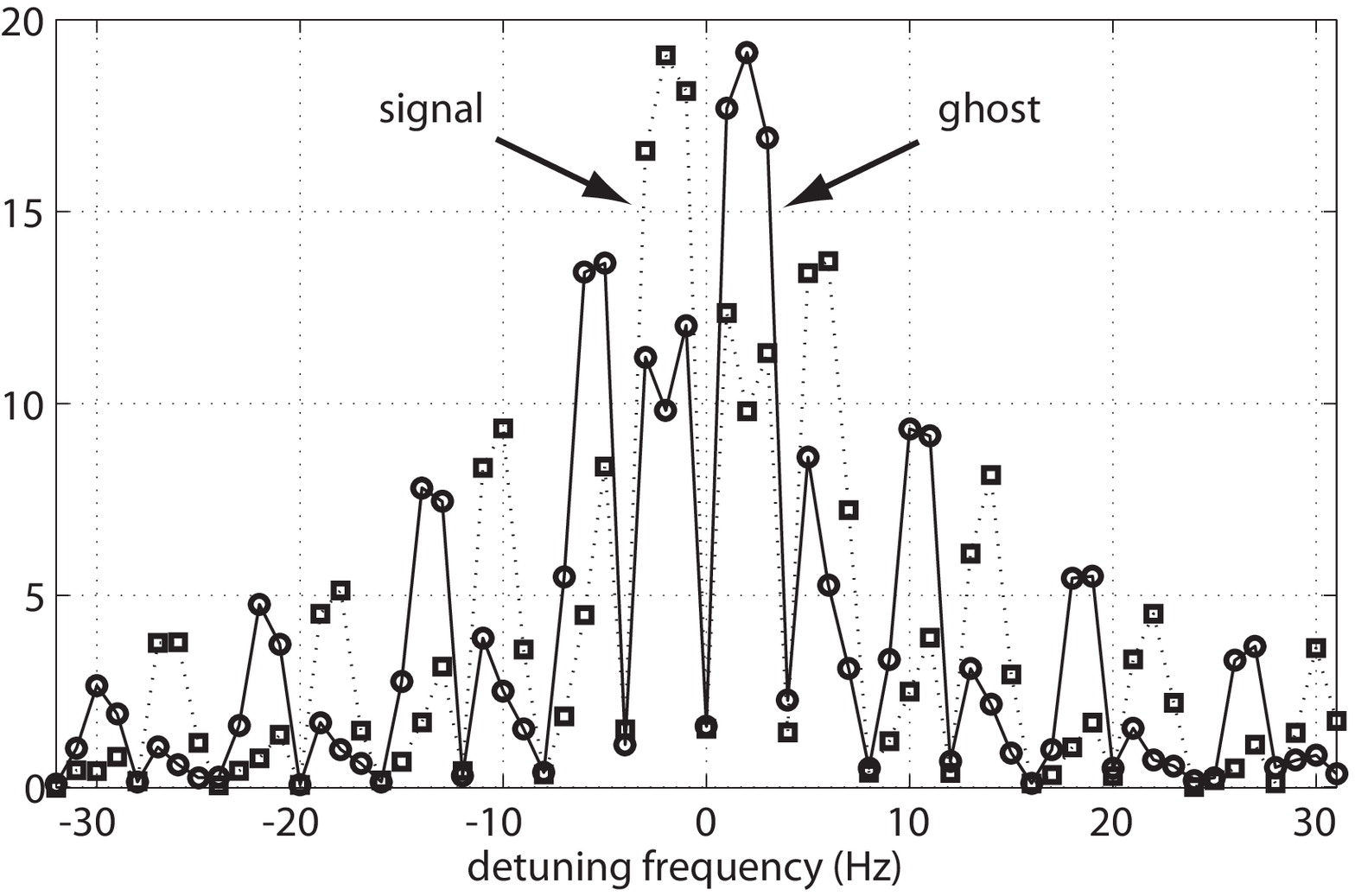}
\caption{Measurement of the temporal frequency instrumental
response. Camera framerate : $\omegaS / 2 \pi = 8$ Hz. Exposure time
: $\TE = 124 \,\rm ms$. $n=4$. Representation of instrumental
responses for the true image (signal) and the twin image (ghost), in
dB. Horizontal axis : detuning frequency $(\omegaLO - \omegaL)/(2
\pi)$, in Hz. Squares : signal (first heterodyne term). Circles :
ghost (second heterodyne term). \label{fig_hetero_BW_mes}}
  \end{center}
\end{figure}

\begin{figure}[h]%[h]
  \begin{center}%
\includegraphics[width = 8.0cm]{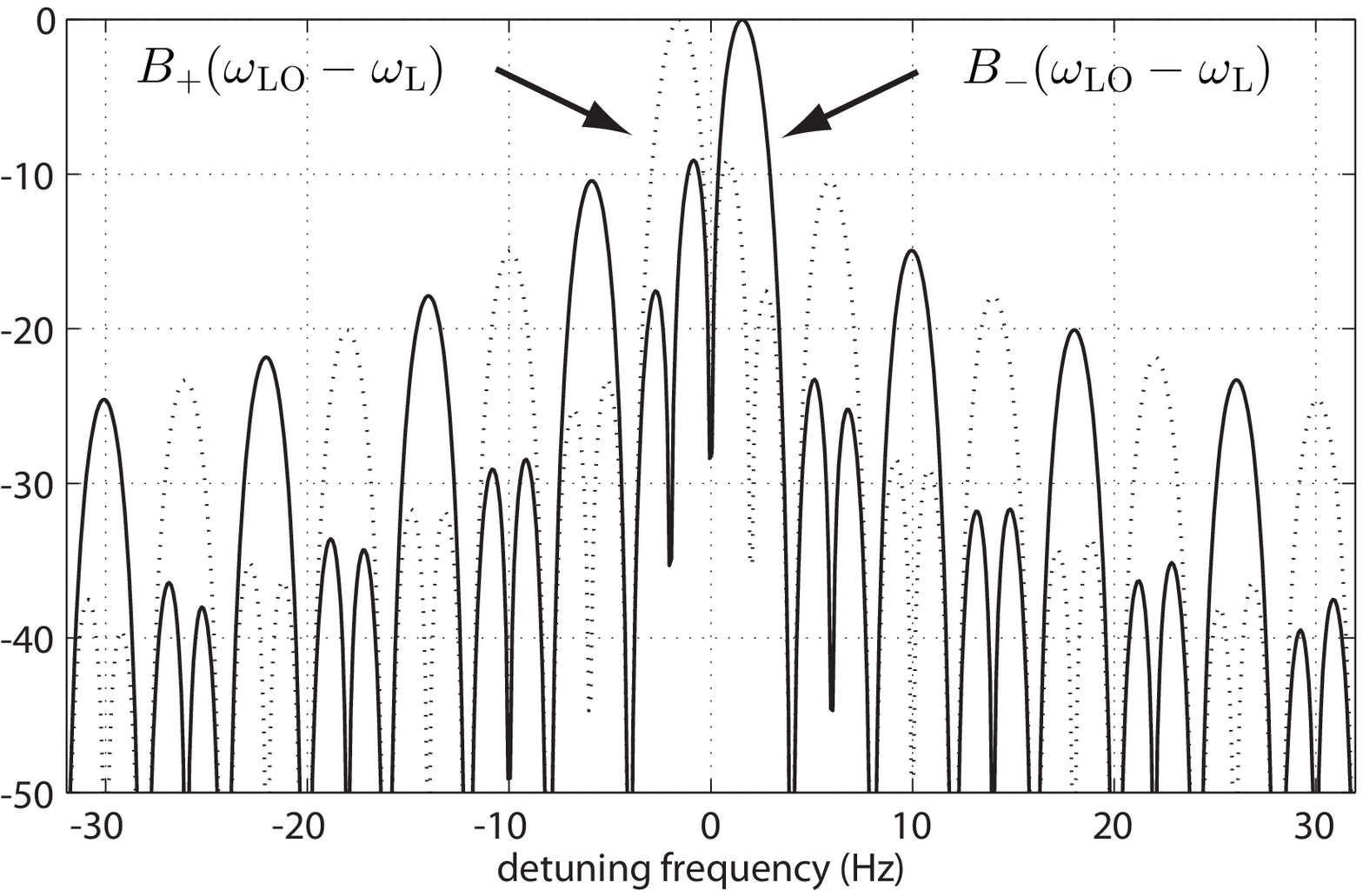}
\caption{Squared amplitude of the instrumental response defined by
eq. \ref{eq_B_omega_defn}. Camera framerate : $\omegaS / 2 \pi = 8$
Hz. Exposure time : $\TE = 124 \,\rm ms$. $n=4$. Representation of
$10 \log_{10} [\left|B_{\pm}(\omegaLO-\omegaL)\right|^2]$, in dB.
Horizontal axis : detuning frequency $(\omegaLO - \omegaL)/(2 \pi)$,
in Hz. Dotted line : $B_+$. Continuous line :
$B_-$\label{fig_hetero_BW}}
  \end{center}
\end{figure}

If we neglect the contribution of the amplitude fluctuations
$\At(t)$ to the laser linewidth, the impulse response functions
for the first and second heterodyne terms ($B _{+}
(\omegaLO-\omegaL)$ and $B _{-} (\omegaLO-\omegaL)$ respectively)
calculated by setting $\Ht(t) = 1$ in expressions
\ref{eq_Itilde1_bis}, \ref{eq_Itilde2_bis} and \ref{eq_F_t} take
the form :
\begin{eqnarray}\label{eq_B_omega_defn}
\nonumber B_{\pm}(\omegaLO-\omegaL) = \sinc
\left((\omegaLO-\omegaL \right) \TE) \\ \times \sum  _{k=1} ^{n}
\, \exp(-2 i k \pi / n) \, \exp (\mp 2 i k \pi \frac{\omegaLO -
\omegaL}{ \omegaS } )
\end{eqnarray}\
The instrumental response $\left| B _\pm (\omegaLO-\omegaL)
\right|^2$ was measured experimentally, by analyzing light
backscattered by a static object. Data points of this instrumental
response are plotted as a function of the detuning frequency in
fig. \ref{fig_hetero_BW_mes}. The square amplitude of the function
described by eq. \ref{eq_B_omega_defn} is represented in fig.
\ref{fig_hetero_BW}. Although their shape is similar, the dynamic
range of the theoretical filter is wider than the measured
response. The difference is attributed to the linewidth
contribution of laser intensity fluctuations. One important thing
to remark about those
instrumental responses is their dissymmetry.\\

$|I_1(\omegaS/n)|^2$ and $|I_2(\omegaS/n)|^2$ are calculated to
represent quantities homogenous to the optical power of the object
field. This power is proportional to the object field power
resulting from the integration of its spectral density in the $B\pm$
windows. We call these distributions \emph{signal} (or \emph {true
image}) and \emph{ghost} (or \emph{twin image}), respectively.
$B\pm$ describe bandpass filters of width $\BWF = \omegaS / n$,
centered on $\omegaLO - \omegaL \pm \omegaS / n$. Consequently, this
scheme allows one to measure the $\omegaLO - \omegaL \pm \omegaS /
n$ frequency components of the object
field fluctuations $\Ht(t)$.\\

\begin{figure}[h]%[h]
  \begin{center}%
   \includegraphics[width = 8.0cm]{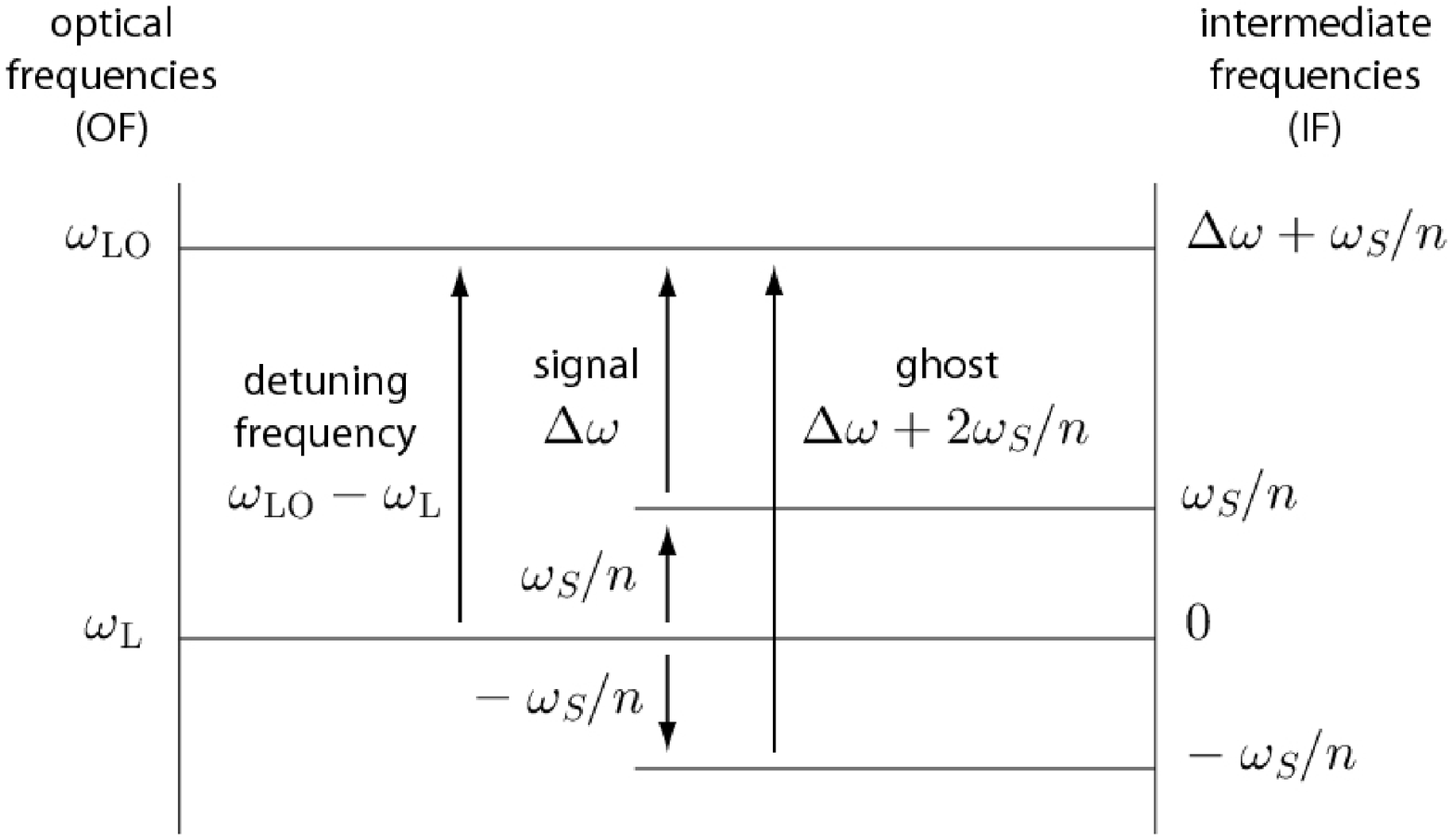}
\caption{Frequency diagram of heterodyne terms spectral components
in the case where the detuning frequency is set to $\omegaLO -
\omegaL =  \Delta \omega - \omegaS/n$.\label{fig_freq_diag}}
  \end{center}
\end{figure}

If we set the detuning frequency to :
\begin{equation}\label{eq_omegaLO_demod}
\omegaLO - \omegaL =  \Delta \omega - \omegaS/n
\end{equation}
the signal will correspond to the $\Delta \omega$ frequency
component of the field and the ghost area to the $\Delta \omega + 2
\omegaS / n$ component. We have
$|I_1(\omegaS/n)|^2 \approx \SONE (\Delta \omega)$ and
$|I_2(\omegaS/n)|^2 \approx \SONE (\Delta \omega +2 \omegaS / n)$.
The corresponding frequency diagram is sketched in
fig. \ref{fig_freq_diag}.

%Spectra presented in Refs. \cite{AtlanGross2006RSI,
%AtlanGrossVitalis2006} were obtained with $\omegaS/(2\pi)$ = 8 Hz
%and $n$ = 4.

\subsection{Homodyne terms. LO and object fields self beating}

In usual heterodyne monodetection schemes, self-beating
contributions are considered as an unavoidable noise component,
which can actually be neglected in a high bandwidth PM detection
scheme. When using low frequency PMs such as camera pixels, these
homodyne contributions (defined by eqs. \ref{eq_I_t3defn} and
\ref{eq_I_t4defn}) have to be taken into account. We show in the
next section how these terms are filtered-out spatially in the
detection process.

\section{Signal demodulation in the spatial domain. Filtering of self-beating contributions}

\subsection{Heterodyne contributions}

The precious advantage of holography is to allow sampling of the
diffracted object complex field (in phase and amplitude), whose
spatial distribution in the detector plane is described by the
$\Hs(x,y)$ function. The time domain demodulation presented in the
previous section enables one to measure the spatial distribution of
a tunable frequency component of the object field, in amplitude and
phase (i.e. both quadratures). Thus, after temporal demodulation,
the actual complex distributions defined by eqs.
\ref{eq_I_xy_hetero1} and \ref{eq_I_xy_hetero2} are
available.\\

As a result of the off-axis configuration, the object-LO field cross
terms (or heterodyne contributions) carry phase factors $\exp \left(
\pm 2 i \pi (x \xi _0 + y \eta _0) \right)$. The relative distance
between each point $(x',y')$ of the object (described by
$\Hs(x',y')$) and the LO focal point $(x'_0,y'_0)$ in the object
plane is encoded in a set of parallel (complex) Young's fringes in
the detector plane $(x,y)$. The periods of these fringes are
$(\lambda d / (x' - x'_0)$ and $\lambda d / (y' - y'_0))$. Their
spatial frequencies are noted $\Delta \xi _0 = (x' - x'_0) /
(\lambda d)$ and $\Delta \eta _0 = (y' - y'_0) / (\lambda d)$.\\

In Fraunhofer conditions, the relative distances between each point
of the object $(x',y')$ and the LO focal point $(x'_0,y'_0)$ are
proportional to the spatial frequencies of the hologram in the
camera plane $(\xi,\eta)$. We have :
\begin{equation}\label{eq_I_xieta_hetero0}
    \left| \Hsf(\xi= x'/(\lambda d),\eta= y'/(\lambda d)) \right|^2
\propto \left| \Hs(x',y') \right|^2
\end{equation}
Hence the requirement for only one spatial Fourier transform to
reconstruct the image \cite{Wagner1999, Schnars2002, Kreis2002} ,
i.e. to calculate the spatial distribution
$\left| \Hs(x',y') \right|^2$.\\

The laser amplitude noise leads to flat-field fluctuations in the
camera plane. The spatial frequencies distribution of this noise is
considered to be a centered dirac :
\begin{eqnarray}\label{eq_flat_field_fluctuations}
\nonumber \int  _{- \Delta x /2} ^{ \Delta x /2} \int  _{- \Delta y
/2} ^{ \Delta y /2}
\left|\As(x,y)\right|^2 \\
\nonumber \times \exp \left(-\frac{2i\pi}{\lambda d}(xx'+yy')\right) \, {\rm d}x \, {\rm d}y \\
\approx \delta(x',y')
\end{eqnarray}
this assumption is implicit in lensless digital holography. The
spatial frequency content of $I_1(x,y)$ , assessed by a FFT, takes
the following form :
\begin{eqnarray}\label{eq_I1_xieta_hetero1}
    \nonumber \left| I_1 (\xi= x'/(\lambda d),\eta=
y'/(\lambda d)) \right|^2 \propto\\
\left| \Hs(x'-x'_0,y'-y'_0) \right|^2
\end{eqnarray}
The ghost distribution is :
\begin{eqnarray}\label{eq_I2_xieta_hetero1}
    \nonumber \left| I_2 (\xi= x'/(\lambda d),\eta=
y'/(\lambda d)) \right|^2 \propto\\
\left| \Hs(x'_0-x',y'_0-y') \right|^2
\end{eqnarray}
In the lensless configuration, both signal and ghost spatial
distributions are focused in the same reconstruction plane. Eqs.
\ref{eq_I1_xieta_hetero1} and \ref{eq_I2_xieta_hetero1} define two
spatial distributions flipped one with respect to the other and
shifted away by $\pm (x'_0,y'_0)$ from the center of the
reconstructed hologram. The spatial instrumental response width
corresponds to less than the distance between two adjacent pixels in
the reconstructed image \cite{Kreis2002}; its contribution is
neglected.

\subsection{Homodyne contributions}

The object and LO self beating contributions $I_3$ and $I_4$ are not
encoded in the spatial complex fringe system which results from the
interference of the object field and the off-axis LO. Under the
assumption of a perfect flat-field amplitude noise (eq.
\ref{eq_flat_field_fluctuations}), the spatial frequency content of
eqs. \ref{eq_I_xy_hetero3} and \ref{eq_I_xy_hetero4} take the form :
\begin{eqnarray}\label{eq_I3_xieta_hetero1}
    I_3 (\xi= x'/(\lambda d),\eta=
y'/(\lambda d)) \propto \delta(x',y')
\end{eqnarray}
and
\begin{eqnarray}\label{eq_I4_xieta_hetero1}
\nonumber I_4 (\xi= x'/(\lambda d),\eta=
y'/(\lambda d)) \propto \\
\Hs(x',y')*\Hs^*(x',y')
\end{eqnarray}
These terms are restituted inline : they are centered on
$(x'=0,y'=0)$ in the hologram reconstructed in the object plane.
This enables one to discriminate them spatially from off-axis
heterodyne contributions.

\section{Experiments}

In previous publications \cite{AtlanGross2006RSI,
AtlanGrossVitalis2006, AtlanGrossLeng2006}, the spatiotemporal
heterodyne detection technique was used to perform parallel imaging.
Here, we will focus our attention on the temporal domain. We will
study in particular how measured frequency spectra are affected by
the camera frame rate, exposure time, and total measurement time.

\subsection{Setup and data acquisition details}

The experimental setup, which is sketched in
Fig.\ref{fig_setup_config}, has been described previously
\cite{GrossDunn2005, AtlanGross2006RSI, AtlanGrossLeng2006,
AtlanGrossVitalis2006}. The light source is a Sanyo DL-7140-201
diode laser ($\lambda=780$ nm, $50$ mW for $95$ mA of current), the
camera is a PCO Pixelfly digital CCD camera ($12$ bit, frame rate
$\omegaS / (2 \pi) \simeq 12.5$ Hz, exposure time $\TE \leq 2 \pi /
\omegaS = 80 ~ \rm{ms}$, with $1280 \times 1024$ pixels of $6.7
\times 6.7 \, \mu \rm m$), and the frequency shifter is a set of two
acousto-optic modulators AOM1 and AOM2 (Crystal Technology;
$\omega_{{\rm AOM}1,2} \simeq 80$ MHz). A neutral density is used to
control the LO beam intensity. The sample is a $1 \textrm{~cm}\times
1 \textrm{~cm}\times 5 \textrm{~cm}$ rectangular PMMA transparent
cell filled with a diluted suspension of latex spheres in water
(Polybead: Polyscience Inc., diameter 0.48 $\mu \rm m$, undiluted
concentration: 2.62$\%$ solids-latex). The Polybead suspension is
diluted by a factor $\simeq 9$ (0.5 ml of undiluted suspension + 4
ml of water). A rectangular aperture ($7 ~\textrm{mm} \times 3 ~
\textrm{cm}$) located just in front of the cell removes the
parasitic light diffused along cell sides. The aperture delimiting
the imaged side of the cell is located at a distance $d \simeq
39$~cm of the camera. This sample is observed in transmission and
the spectrum of the light dynamically scattered by the latex spheres
in brownian motion is measured by sweeping the AOM1 frequency so
that the detuning frequency $\omegaLO - \omegaL$ is swept from $0$
to $16$ kHz in 81 frequency points (200 Hz step). For each frequency
point a sequence of 32 consecutive CCD frames is recorded to the PC
computer hard disk. For each frequency sweep, $32\times 81=2592$
images are recorded.

\subsection{Data analysis}

A first experiment consists of acquiring data with 3 frequency
sweeps with $\TE = 80$ ms, $\TE = 20$ ms and $\TE = 5$ ms. For each
frequency point, a $n=4$ phase demodulation is performed by using
either 4, 8, 16 or the complete set of 32 recorded images. When more
than 4 images are used, a first $n = 4$ phase demodulation is done
with images 1 to 4, which is followed by a second demodulation with
images 5 to 8, and so on. The complex signal resulting from the
demodulation of each set of 4 images is then summed to get the final
demodulation result. The total measurement time of one frequency
point vary from 320 ms ($4 \times 80$ ms) to 2.6 s ($32 \times 80$
ms), while the total exposure time vary from 20 ms ($4 \times \TE$
with $\TE = 5$ ms)
to 2.6 s ($32 \times \TE $ with $\TE = 80$ ms).\\

\begin{figure}[]%[h]
  \begin{center}%
\includegraphics[width = 4.0cm]{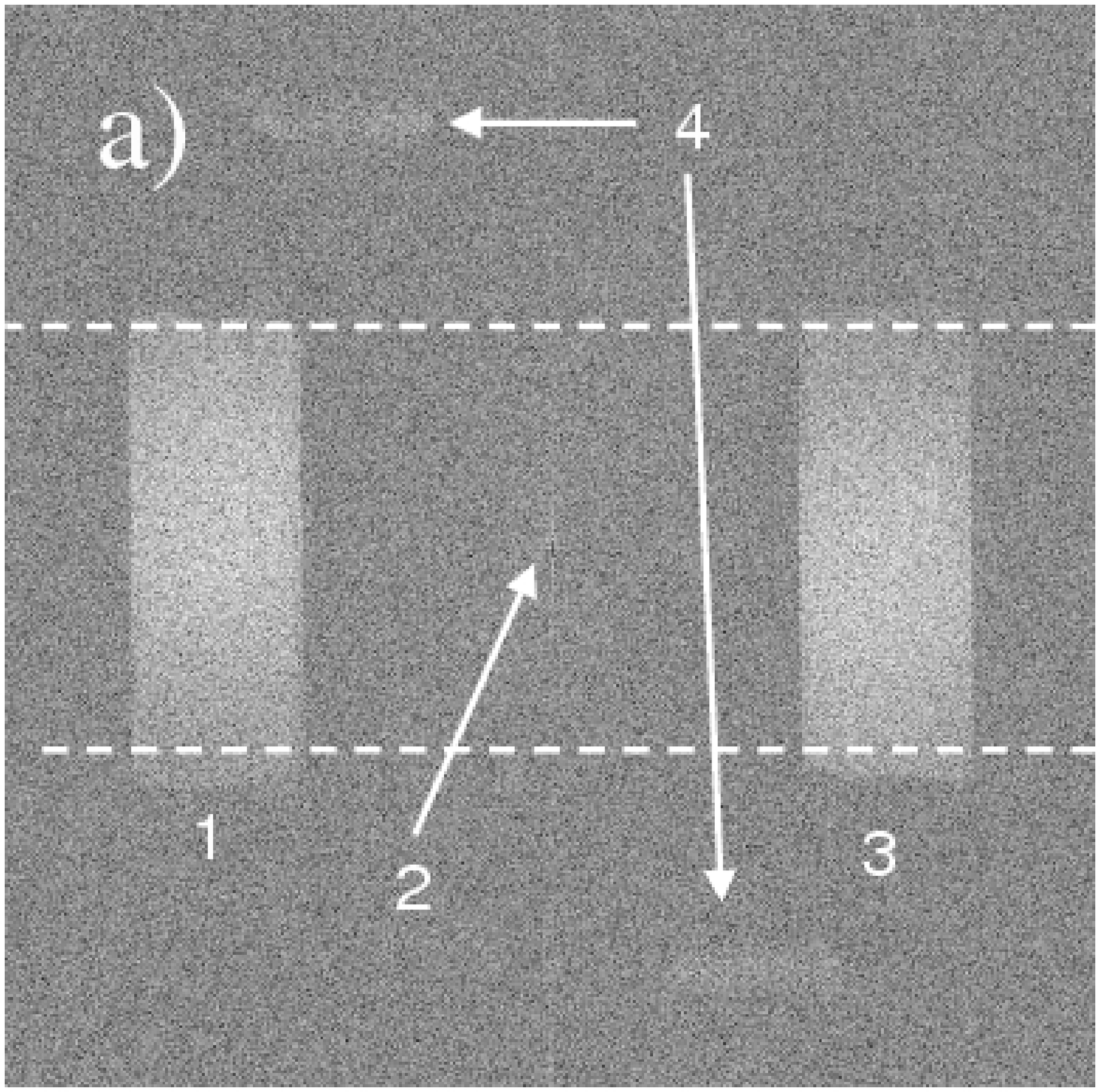}
\includegraphics[width = 4.0cm]{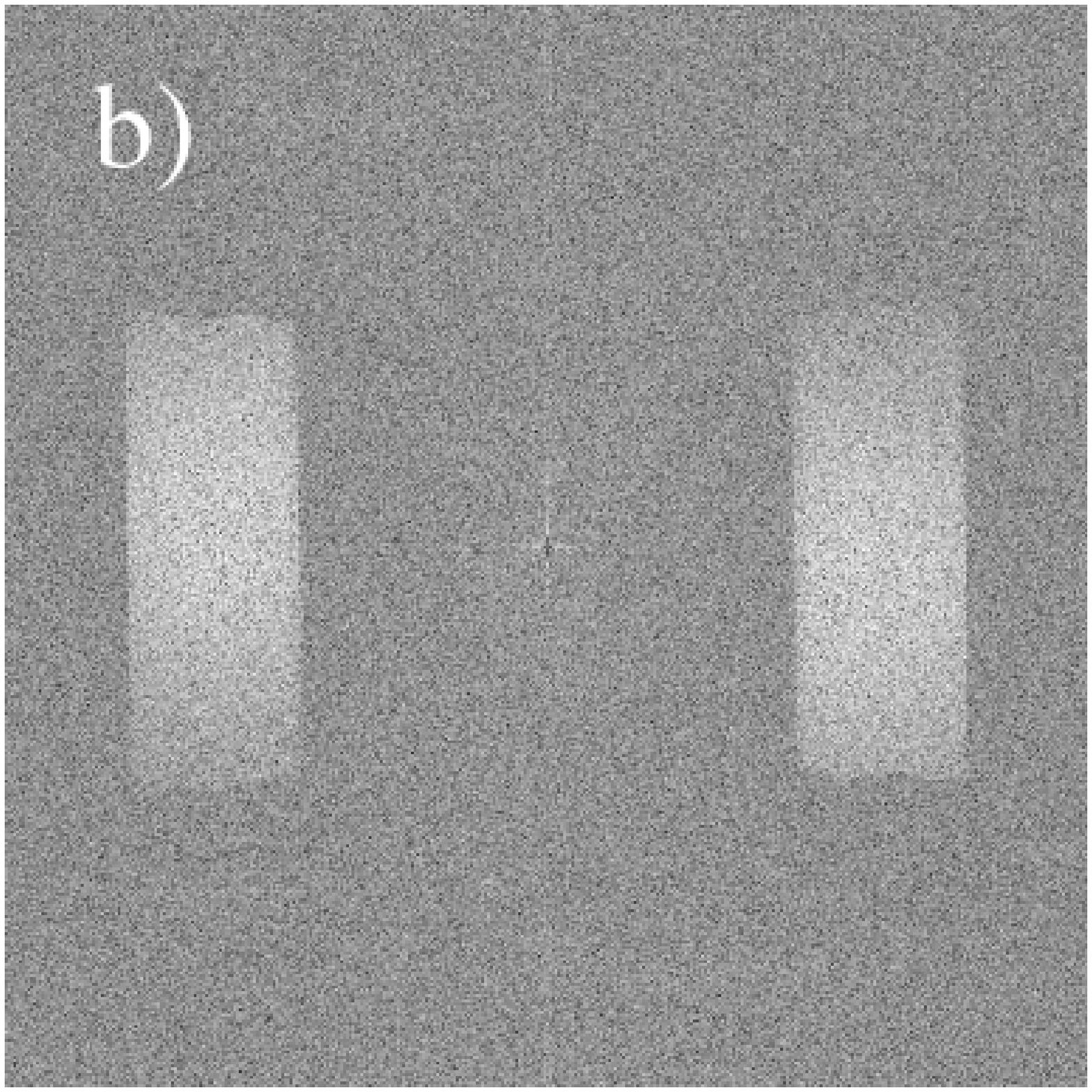}
\includegraphics[width = 4.0cm]{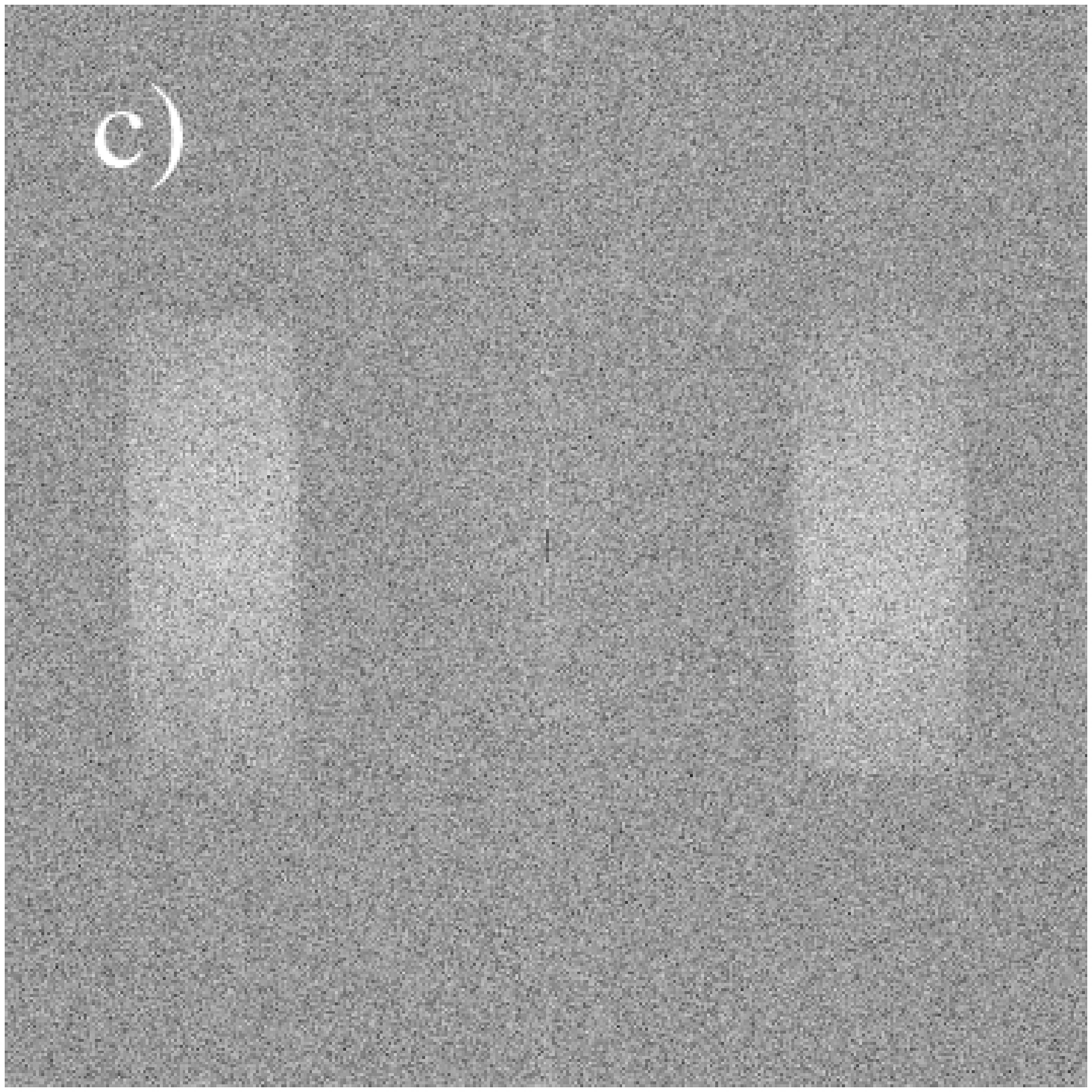}
\includegraphics[width = 4.0cm]{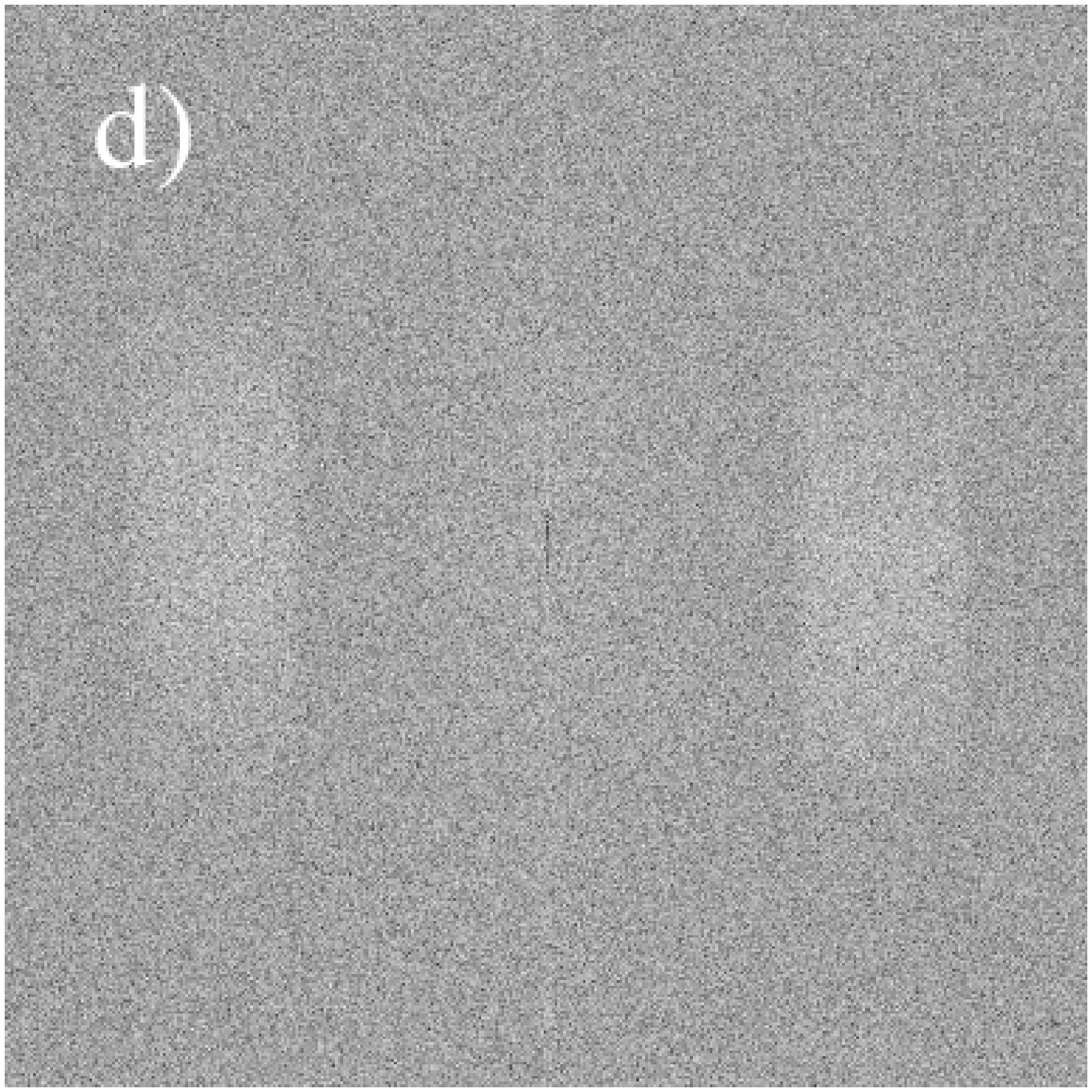}
\caption{Images ($1024 \times 1024$ pixels) of the sample for
$(\omegaLO - \omegaL)/(2 \pi)$ equal to 0 Hz (a), 400 Hz (b), 4000
Hz (c) and 8000 Hz (80 ms image exposure time,  4-image
demodulation. Arbitrary logarithmic scale display.}
\label{fig_images_80ms_4im}
  \end{center}
\end{figure}

Fig. \ref{fig_images_80ms_4im} represents the resulting images of
the rectangular aperture, illuminated in transmission via the
diffusing cell, for an exposure time $\TE =$ 80 ms, and for a
detuning frequency $(\omegaLO - \omegaL) / (2 \pi)$ of 0 Hz (a), 400
Hz (b), 4000 Hz (c) and 8000 Hz (d). The images are k-space images
obtained by FFT of the hologram measured in the CCD plane and $n=4$
phase demodulation with 4 images per frequency point. They are
displayed in arbitrary logarithmic scale. The true image (signal) of
the rectangular aperture is the region 1 in fig.
\ref{fig_images_80ms_4im}(a). This distribution is the spectral
component of the object field of frequency $\omegaL - \omegaS / n$.
The twin image (ghost) of the aperture is the region 3. In
accordance with eq. \ref{eq_I1_xieta_hetero1} and
\ref{eq_I2_xieta_hetero1}, it is symmetric  with respect to the
center (tag 2) of the k-space plane (null spatial frequency). The
twin image corresponds to the $\omegaL + \omegaS/n$ spectral
component. Increasing the detuning frequency $\omegaLO - \omegaL$,
the true and twin images of the aperture become darker and darker.
Nevertheless, they are still visible for a 8000 Hz offset. One can
notice that for null detuning (fig. \ref{fig_images_80ms_4im}(a))
some statically scattered parasitic light is detected out of the
aperture image (region 4). This parasitic light is no more detected
when the frequency offset is non zero (images (b) to
(d)).\\

\begin{figure}[]%[h]
  \begin{center}%
\includegraphics[width = 8.0cm]{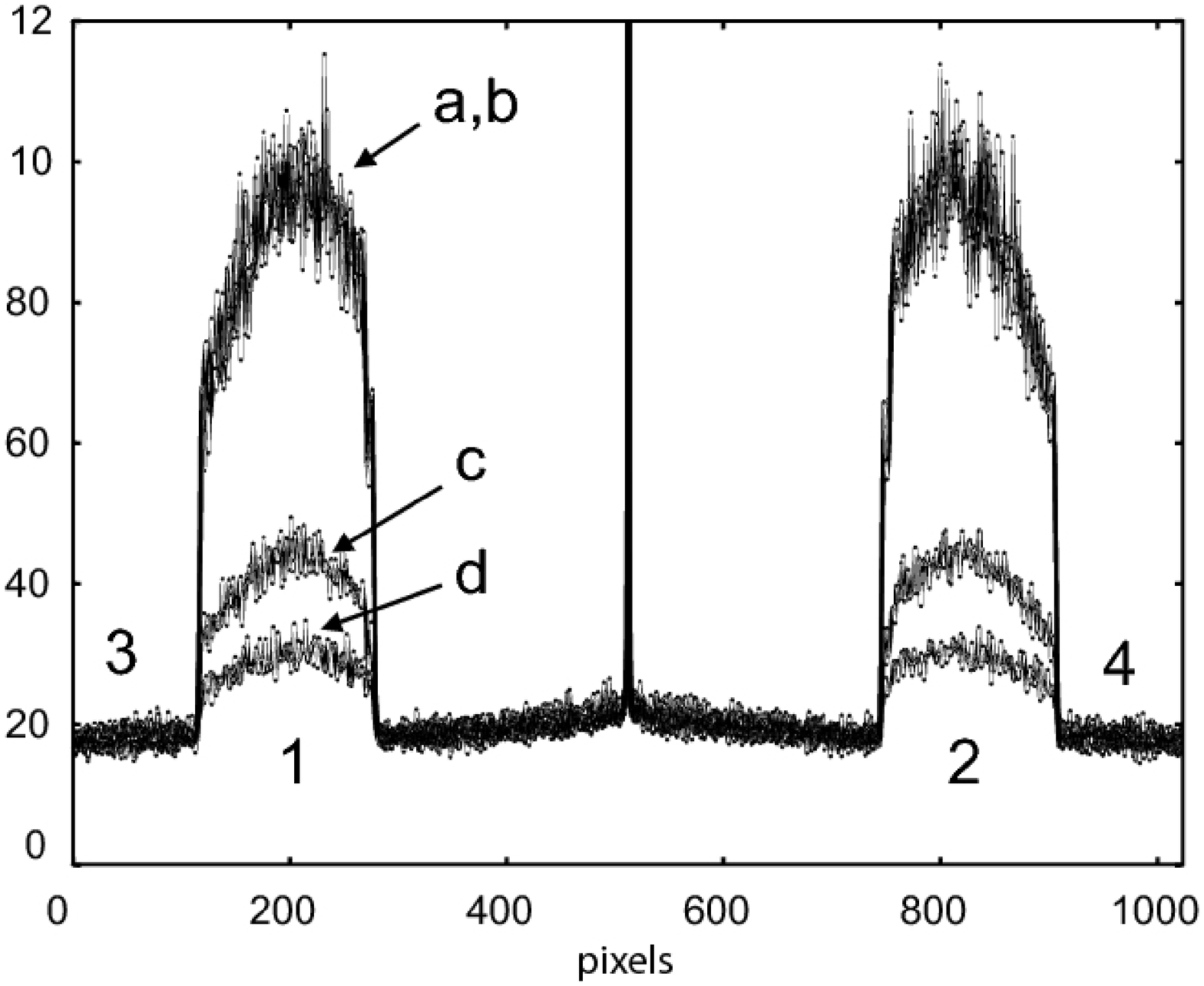}
\caption{Traces obtained by summation along columns of
Fig.\ref{fig_images_80ms_4im}(a) to (d) intensities. Curves a to d
correspond to a detuning frequency $(\omegaLO - \omegaL) / (2 \pi)$
equal to 0 Hz (a), 400 Hz (b), 4000 Hz (c) and 8000 Hz (d).
Horizontal scale is the image horizontal pixel index. Vertical scale
is in linear arbitrary units.} \label{fig_curve_80ms_4im}
  \end{center}
\end{figure}

To  perform a quantitative analysis of the signal, we have summed
the signal intensity along  the columns of the images represented
on Fig.\ref{fig_images_80ms_4im}, from line 312 to line 712
(horizontal dashed lines  on Fig.\ref{fig_images_80ms_4im}(a)).
These traces are represented on fig. \ref{fig_curve_80ms_4im}. The
true image corresponds to region 1, the twin image to region 2.
Regions 3 and 4 correspond to the background signal. As mentioned
in \cite{GrossAl-Koussa2003, GrossDunn2005}, this background
signal is due to the shot-noise on the LO. Taking into account the
heterodyne gain, the background signal corresponds to an optical
signal of one photo electron per pixel for the whole measurement
sequence. The background signal provides here a very simple
absolute calibration for the optical signal diffused by the
sample. Here, in the center of the aperture, the sample diffused
signal is about 3.5 times the background i.e. 3.5 photo electron
per pixel.

\subsection{Spectra of Doppler-broadened light in the kHz range}

\begin{figure}[]%[h]
  \begin{center}%
\includegraphics[width = 8.0cm]{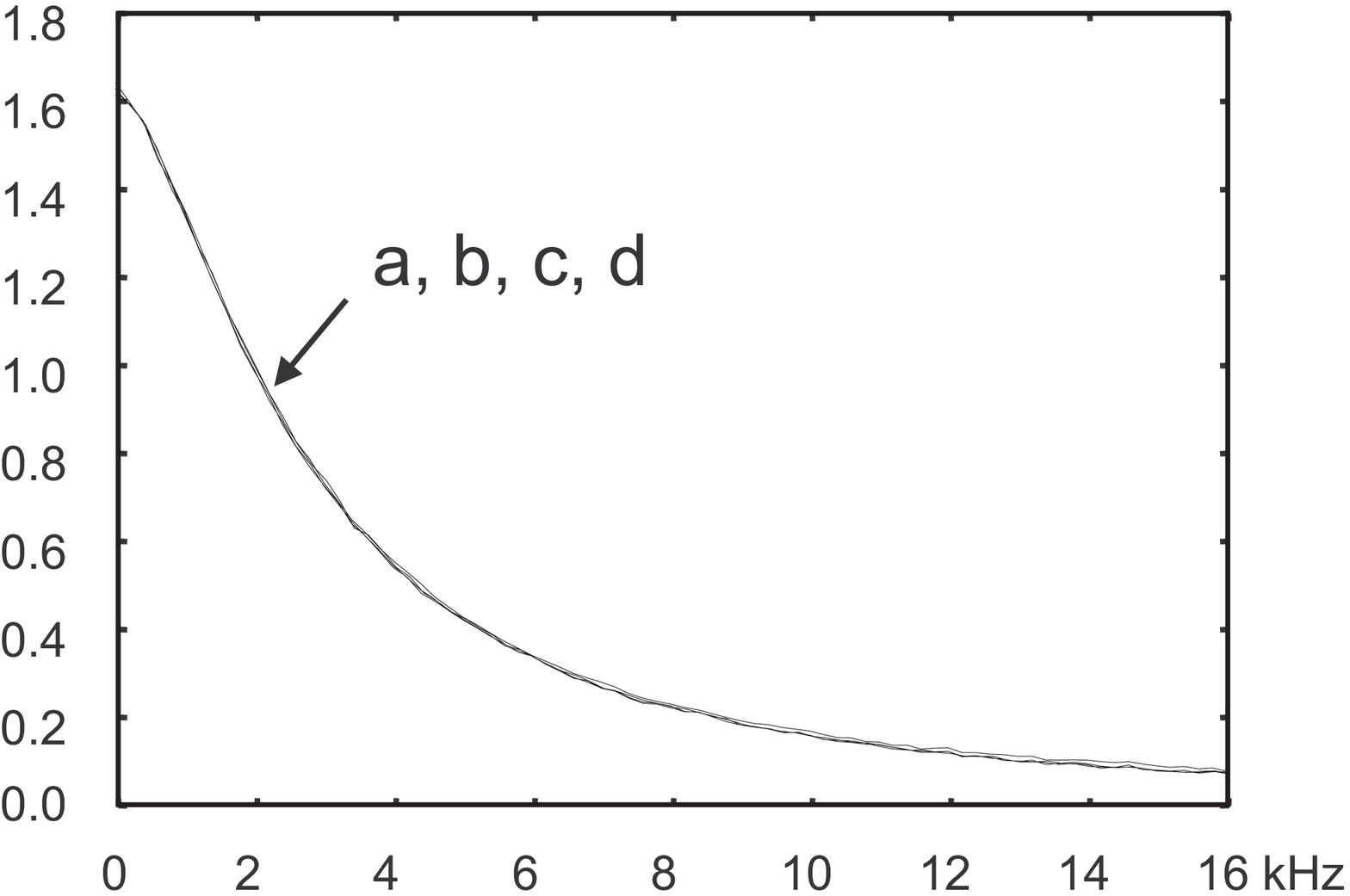}
\caption{Frequency spectra of the light diffused through a
suspension of latex particles in brownian motion. Exposure time is
$\TE=80$ ms. Demodulation is performed with 4 (a), 8 (b), 16 (c) and
32 (d) images. Horizontal axis is the detuning frequency $(\omegaLO
- \omegaL)/(2\pi)$ in kHz. Vertical scale is in linear arbitrary
units. The four curves overlap.} \label{fig_spectrum_80ms}
  \end{center}
\end{figure}

\begin{figure}[]%[h]
  \begin{center}%
\includegraphics[width = 8.0cm]{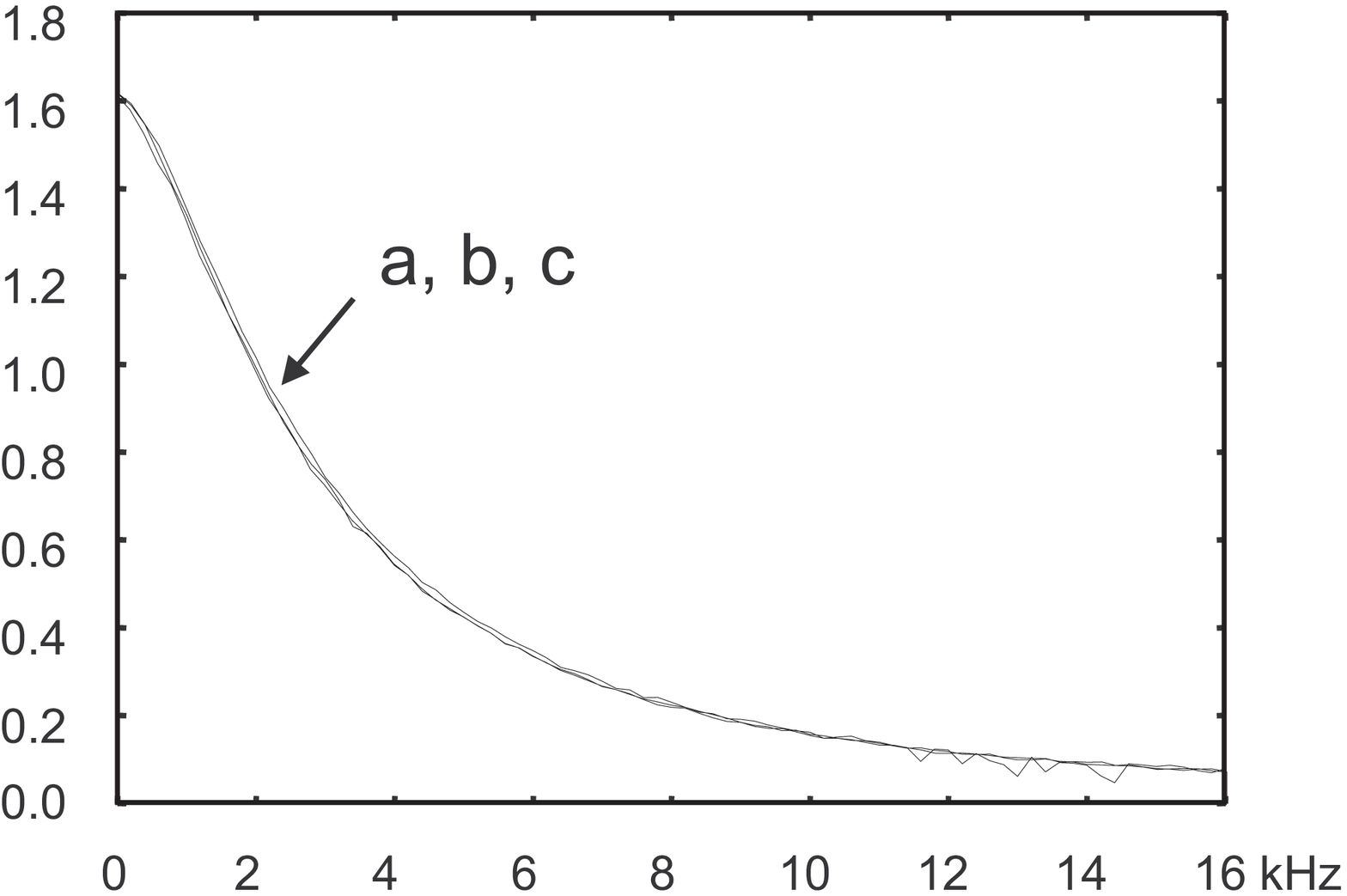}
\caption{Spectra measured with exposure time $\TE=80$ ms (a), 20 ms
(b), 5 ms (c), and 4-image demodulation. Note that curve (a) the
same as Fig.\ref{fig_spectrum_80ms} (4 images and 80 ms). Vertical
scale is in linear arbitrary units. The three curves overlap.}
\label{fig_spectrum_80_5ms}
  \end{center}
\end{figure}

From the reconstructed images (Fig.\ref{fig_images_80ms_4im}), we
have calculated the shape of the first order spectrum of the
diffused object field. To get these spectra, we have calculated the
sum of the area of the true image and twin image areas (regions 1
and 2 in the traces of fig. \ref{fig_curve_80ms_4im}), and
subtracted the background (regions 3 and 4 in fig.
\ref{fig_curve_80ms_4im}). Fig. \ref{fig_spectrum_80ms} and Fig.
\ref{fig_spectrum_80_5ms} show brownian frequency spectra of the
light dynamically scattered through the cell. These curves have been
normalized by the area under the raw lineshapes.\\

Data was collected for detuning frequencies up to 16 kHz, much
larger than the heterodyne receiver bandwidth (0.4 Hz for curve (d),
reciprocal of the 2.6 s measurement time). The shape of the spectrum
does not depend on the number of images used to perform the
demodulation, as we can see in Fig. \ref{fig_spectrum_80ms}. In the
range of exposure times $\TE$ (of one frame) used for the
measurement, the shape of the spectra reported on fig.
\ref{fig_spectrum_80_5ms} does vary with $\TE$ for the following
values of the latter : $5$ ms (curve c), 20 ms (curve b), and 80 ms
(curve a). Fig.\ref{fig_spectrum_80ms} and
Fig.\ref{fig_spectrum_80_5ms}, which exhibit the same lineshape,
show that these spectra do not depend on the total exposure time,
which varies from 20 ms (Fig. \ref{fig_spectrum_80_5ms}(c)) to 2.6 s
(Fig. \ref{fig_spectrum_80ms}(d)). One must notice that these
results are valid because the width of the spectrum ($\simeq 2.5$
kHz half width) is large compared to the instrumental response. In
the all the presented results, the instrumental response is always
smaller than $1 / \TE \leq 200 $ Hz (width of the $\sinc$ factor in
Eq.\ref{eq_B_omega_defn}).\\

\begin{figure}[]%[h]
  \begin{center}%
\includegraphics[width = 8.0cm]{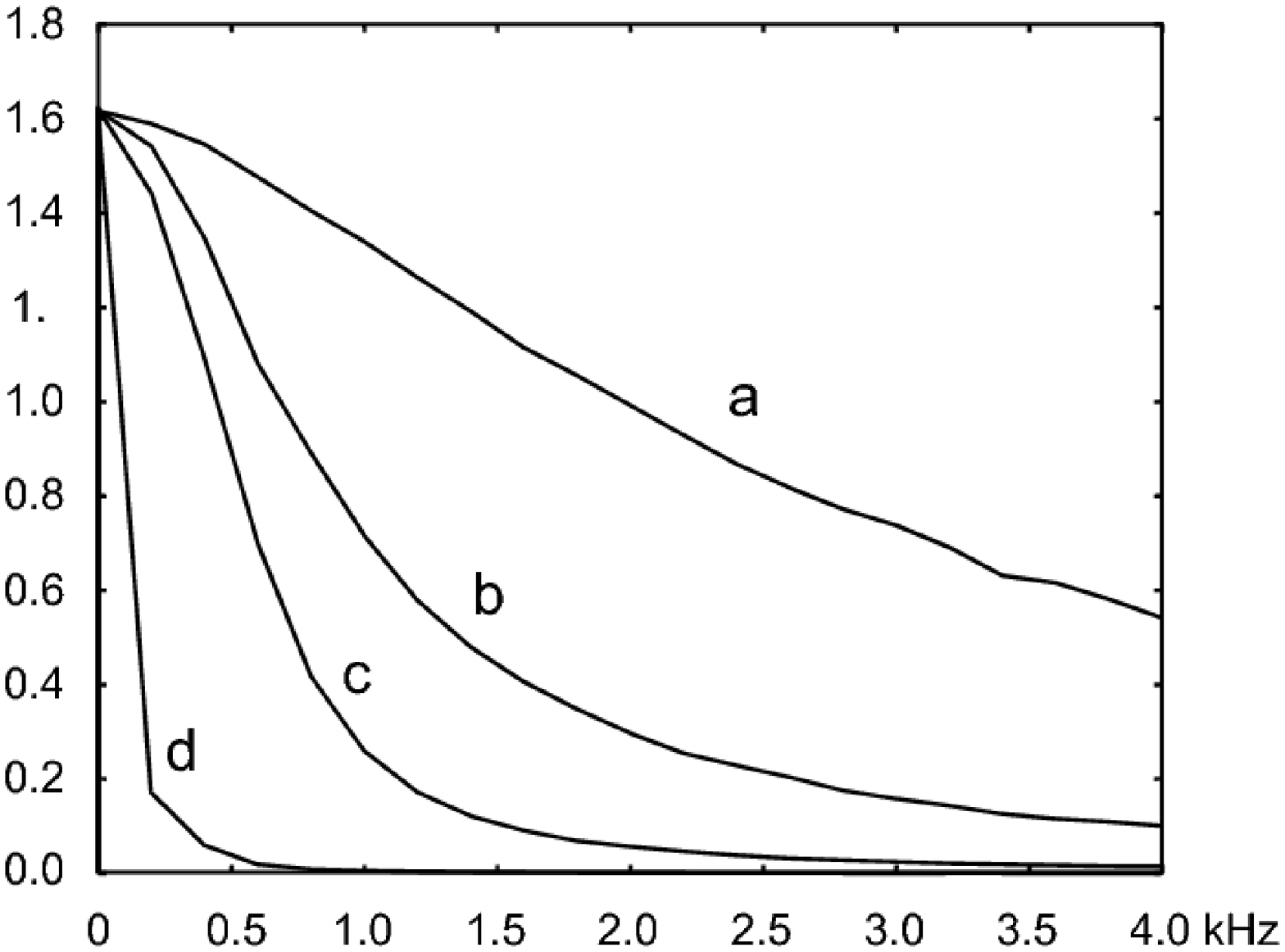}
\caption{Frequency lineshapes of the light diffused through the cell
for different concentrations of latex spheres. Exposure time is $\TE
= 80$ ms. 4-image demodulation. Horizontal axis is the detuning
frequency $(\omegaLO - \omegaL)/(2\pi)$ in kHz. Vertical scale is in
linear arbitrary units. Volumic concentration of latex beads : $2.9
\times 10^{-3}$ (a), $1.5 \times 10^{-3}$ (b), $7.3 \times 10^{-4}$
(c), to $3.6 \times 10^{-4}$ (d).} \label{fig_spectrum_80m_dilues}
  \end{center}
\end{figure}

Additionally, we measured the first order spectrum of the object
field in fixed instrumental conditions (4-phase demodulation,
exposure time $\TE$ = 80 ms) to study different samples. The
scattering parameters of the suspension were changed in the
following manner : the latex beads concentration was halved from one
measurement to another. The four Doppler lineshapes reported on fig.
\ref{fig_spectrum_80m_dilues} correspond to latex volumic fractions
ranging from $2.9 \times 10^{-3}$ (curve a) to $3.6 \times 10^{-4}$
(curve d).

\section{Discussion}

The presented technique uses and goes beyond the concept of
time-average holography. In time-averaged holography, the detector,
whose exposure time $\TE$ is large, behaves as a low-pass filter of
bandwidth $1/\TE$. It selects the field components $\EO(\omega)$
whose frequencies $\omega = \omegaL + \Delta \omega$ are close to
$\omegaLO$, i.e. which satisfy $|\Delta \omega| \TE < 1$). This
low-pass filtering effect is used to study vibrating objects like
musical instruments \cite{DemoliDemoli2005} by imaging the regions
which are not moving or which move with a given phase and velocity.
In the first case (selection of the non moving regions), the LO beam
is not shifted in frequency \cite{Powell1965, Picart2003}. In the
second case (moving regions), the LO beam is modulated at the object
vibration frequency \cite{Aleksoff1969, Lokberg1984} (which, in the
temporal frequency domain, corresponds to the generation of adequate
LO frequency sidebands.\\

Our scheme enables detection of a tunable frequency component of the
light with a sharp bandwidth defined by the inverse of the
acquisition time of a sequence of $n$ images. But reading-out an
optical beat at typically low camera frame rates is difficult in
practice in the case of a weak-amplitude object field, because the
low frequency part of the temporal frequency spectrum is highly
noisy : it contains LO and object field self-beating contributions.
To discriminate the signal from these noise contributions, a
lensless Fourier off-axis configuration is used to encode, in a
Young fringe system, the distribution of relative positions of
object points with respect to the reference point. This scheme
allows to restitute the object field spatial distribution in the
Fourier reciprocal space of the detection plane. Having recourse to
an off-axis configuration to sample the heterodyne optical beat
enables spatial frequency discrimination of the \emph{heterodyne}
terms (object and reference light cross terms) carrying useful
information from the object and reference light self-beating terms
carrying what turns out to be unwanted information. Time domain and
spatial domain modulation/demodulation schemes present strong
similarities. Equation \ref{eq_I_xy_hetero1} is the spatial
counterpart of eq. \ref{eq_I1}. Signal and ghost distributions are
beating at $\pm \omegaS / n$ in time domain. This beating enables to
perform a $n$-phase temporal demodulation of the recorded image
sequence to assess object and ghost fields in quadrature. Since
these complex fields are spatially modulated because of the off-axis
interferometry configuration (according to eqs.
\ref{eq_I_xy_hetero1} and \ref{eq_I_xy_hetero2}), their
distributions in the object plane are translated by $\pm(x'_0,y'_0)$
(respectively) away from $(0,0)$-centered homodyne contributions.
This is highly valuable since it allows one to filter them out
spatially.

\section{Conclusion}

The presented scheme, based on heterodyne optical mixing onto a
parallel detector used as a selective frequency filter, is
particulary suited to high resolution spectral imaging. An tunable
frequency component of the object field is acquired at a time. The
camera finite exposure time and the finite measurement time lead to
a frequency domain bandpass filter, which width is the reciprocal of
the measurement time. The small dissymmetry of the frequency domain
instrumental response should be taken into account in quantitative
measurements. The available range of frequency shifts at which
measurements can be made is not limited to the photo-mixer bandwidth
(contrary to wide-field laser Doppler \cite{Serov2005}), and
benefits from heterodyne amplification. The measurement is sensitive
: thanks to the off-axis holographic setup, the self beating
intensity contributions resulting from laser instabilities and
scattered light self interference within the camera bandwidth can be
efficiently filtered-out. The propensity to separate spatially cross
terms intensity contributions from self beating contributions lies
in the use of a spatial heterodyne method. The presented scheme is
the association of a spatial and a
temporal heterodyne detection.\\

Both temporal and spatial resolution are potentially high. This
combination is enabled by a wide field measurement performed at one
frequency point at a time. The data transfer rate bottleneck implies
a tradeoff between the number of pixels in the image and the frame
rate of the detector, which should be guided by the application
needs in terms of temporal, spectral and spatial resolution. In a
few words, combining a spatiotemporal modulation and demodulation of
a coherent probe light allows one to achieve a sensitive wide field
detection of a tunable Hertz-resolved spectral component with an
array detector.\\

The authors acknowledge support from the French National Research
Agency (ANR) and from Paris VI University (BQR grant).\\

%Corresponding author: atlan@lkb.ens.fr

%\bibliographystyle{apsrev}%pas de titres; format compact
%\bibliographystyle{unsrt}
%\bibliography{E:/LaTeX/bibliographies/Biblio_Atlan_060908}

\begin{thebibliography}{10}

\bibitem{Forrester1961}
A.~T. Forrester.
\newblock Photoelectric mixing as a spectroscopic tool.
\newblock {\em J. Opt. Soc. Am.}, 51:253, 1961.

\bibitem{bk_BernePecora2000}
B.~J. Berne and R.~Pecora.
\newblock {\em Dynamic Light Scattering}.
\newblock Dover, 2000.

\bibitem{Brown1983}
Judith~C. Brown.
\newblock Optical correlations and spectra.
\newblock {\em American Journal of Physics}, 51(11):1008--1011, 1983.

\bibitem{Chung1997}
DS~Chung, KY~Lee, and E~Mazur.
\newblock Fourier-transform heterodyne spectroscopy of liquid and solid
  surfaces.
\newblock {\em Applied physics. B, Lasers and optics}, 64:1, 1997.

\bibitem{Stern1977}
MD~Stern, DL~Lappe, PD~Bowen, JE~Chimosky, GA~Holloway, HR~Keiser,
and
  RL~Bowman.
\newblock Continuous measurement of tissue blood flow by laser-doppler
  spectroscopy.
\newblock {\em American journal of physiology}, 232(4):H441, 1977.

\bibitem{EssexByrne1991}
TJH Essex and PO~Byrne.
\newblock A laser doppler scanner for imaging blood flow in skin.
\newblock {\em J Biomed Eng}, 13(3):189, 1991.

\bibitem{Briers2001}
J.~D. Briers.
\newblock Laser doppler, speckle and related techniques for blood perfusion
  mapping and imaging.
\newblock {\em Physiological Measurement}, 22:R35--R66, 2001.

\bibitem{GrossAl-Koussa2003}
M.~Gross, P.~Goy, and M.~Al-Koussa.
\newblock Shot-noise detection of ultrasound-tagged photons in
  ultrasound-modulated optical imaging.
\newblock {\em Optics Letters}, 28:2482--2484, 2003.

\bibitem{Atlan2005}
M.~Atlan, B.C. Forget, F.~Ramaz, A.C. Boccara, and M.~Gross.
\newblock Pulsed acousto-optic imaging in dynamic scattering media with
  heterodyne parallel speckle detection.
\newblock {\em Opt. Lett.}, 30(11):1360--1362, 2005.

\bibitem{GrossDunn2005}
M.~Gross, P.~Goy, B.C. Forget, M.~Atlan, F.~Ramaz, A.C. Boccara, and
A.K. Dunn.
\newblock Heterodyne detection of multiply scattered monochromatic light with a
  multipixel detector.
\newblock {\em Opt. Lett.}, 30(11):1357--1359, 2005.

\bibitem{AtlanGross2006RSI}
M.~Atlan and M.~Gross.
\newblock Laser doppler imaging, revisited.
\newblock {\em Review of Scientific Instruments}, 77(11), 2006.

\bibitem{AtlanGrossVitalis2006}
M.~Atlan, M.~Gross, T.~Vitalis, A.~Rancillac, B.~C. Forget, and
A.~K. Dunn.
\newblock Frequency-domain, wide-field laser doppler in vivo imaging.
\newblock {\em Optics Letters}, 31(18), 2006.

\bibitem{AtlanGrossLeng2006}
M.~Atlan, M.~Gross, and J.~Leng.
\newblock Laser doppler imaging of microflow.
\newblock {\em Journal of the European Optical Society - Rapid publications},
  1:06025--1, 2006.

\bibitem{LesaffreAtlan2006}
Max Lesaffre, Michael Atlan, and Michel Gross.
\newblock Effect of the photon's brownian doppler shift on the
  weak-localization coherent-backscattering cone.
\newblock {\em Physical Review Letters}, 97(3):033901, 2006.

\bibitem{Siegman1966}
A.E. Siegman.
\newblock The antenna properties of optical heterodyne receivers.
\newblock {\em Applied Optics}, 5(10):1588, 1966.

\bibitem{Gabor1948}
D.~Gabor.
\newblock A new microscopic principle.
\newblock {\em Nature}, 161:777--778, 1948.

\bibitem{Goodman1967}
J.~W. Goodman and R.~W. Lawrence.
\newblock Digital image formation from electronically detected holograms.
\newblock {\em Applied Physics Letters}, 11(3):77--79, 1967.

\bibitem{Stroke1965}
George~W. Stroke.
\newblock Lensless fourier-transform method for optical holography.
\newblock {\em Applied Physics Letters}, 6(10):201--203, 1965.

\bibitem{Wagner1999}
Christoph Wagner, Sonke Seebacher, Wolfgang Osten, and Werner
Juptner.
\newblock Digital recording and numerical reconstruction of lensless fourier
  holograms in optical metrology.
\newblock {\em Applied Optics}, 38:4812--4820, 1999.

\bibitem{Schnars1994}
U.~Schnars.
\newblock Direct phase determination in hologram interferometry with use of
  digitally recorded holograms.
\newblock {\em Journal of Optical Society of America A.}, 11(7):2011, 1994.

\bibitem{Schnars2002}
U.~Schnars and W.~P.~O. Juptner.
\newblock Digital recording and numerical reconstruction of holograms.
\newblock {\em Meas. Sci. Technol.}, 13:R85--R101, 2002.

\bibitem{LeClercGross2000}
F.~LeClerc, L.~Collot, and M.~Gross.
\newblock Numerical heterodyne holography with two-dimensional photodetector
  arrays.
\newblock {\em Optics Letters}, 25(10):716--718, 2000.

\bibitem{Creath1985}
Katherine Creath.
\newblock Phase-shifting speckle interferometry.
\newblock {\em Applied Optics}, 24(18):3053, 1985.

\bibitem{Yamaguchi1997}
I.~Yamaguchi and T.~Zhang.
\newblock Phase-shifting digital holography.
\newblock {\em Optics Letters}, 18:31, 1997.

\bibitem{Indebetouw1999}
Guy Indebetouw and Prapong Klysubun.
\newblock Space--time digital holography: A three-dimensional microscopic
  imaging scheme with an arbitrary degree of spatial coherence.
\newblock {\em Applied Physics Letters}, 75(14):2017--2019, 1999.

\bibitem{Indebetouw2001}
G.~Indebetouw and P.~Klysubun.
\newblock Spatiotemporal digital microholography.
\newblock {\em Optical Society of America Journal A}, 18:319--325, February
  2001.

\bibitem{Kreis2002}
Thomas~M. Kreis.
\newblock Frequency analysis of digital holography.
\newblock {\em Optical Engineering}, 41(4):771--778, 2002.

\bibitem{DemoliDemoli2005}
N.~Demoli and I.~Demoli.
\newblock Dynamic modal characterization of musical instruments using digital
  holography.
\newblock {\em Opt. Express}, 13:4812--4817, 2005.

\bibitem{Picart2003}
Pascal Picart, Eric Moisson, and Denis Mounier.
\newblock Twin-sensitivity measurement by spatial multiplexing of digitally
  recorded holograms.
\newblock {\em Applied Optics}, 42(11):1947--1957, 2003.

\bibitem{Powell1965}
R.~L. Powell and K.~A. Stetson.
\newblock Interferometric vibration analysis by wavefront reconstruction.
\newblock {\em J. Opt. Soc. Am.}, 55:1593, 1965.

\bibitem{Aleksoff1969}
C.~C. Aleksoff.
\newblock Time average holography extended.
\newblock {\em Appl. Phys. Lett}, 14:23, 1969.

\bibitem{Lokberg1984}
O.~J. Lokberg.
\newblock Espi- the ultimate holographic tool for vibration analysis.
\newblock {\em J. Acoust. Soc. Am.}, 55:1783, 1984.

\bibitem{Serov2005}
A.~Serov, B.~Steinacher, and T.~Lasser.
\newblock Full-field laser doppler perfusion imaging monitoring with an
  intelligent cmos camera.
\newblock {\em Opt. Ex.}, 13(10):3681, 2005.

\end{thebibliography}

\pagebreak
\section{List of figure captions}
\begin{enumerate}

  \item Coherent spectral detection schemes : homodyne (a) and
heterodyne (b) optical mixing. PM : photomixer (square-law
detector). SA : spectrum analyzer. $\EO$ object field. $\ELO$ :
local oscillator field. $\SONE$, $\STWO$ : first and second order
object field spectral distributions.

  \item Off-axis lensless Fourier configuration for heterodyne
holography. $L$ : laser. $M$ : mirror. $BS$ : beam splitter.

  \item Measurement of the temporal frequency instrumental
response. Camera framerate : $\omegaS / 2 \pi = 8$ Hz. Exposure time
: $\TE = 124 \,\rm ms$. $n=4$. Representation of instrumental
responses for the true image (signal) and the twin image (ghost), in
dB. Horizontal axis : detuning frequency $(\omegaLO - \omegaL)/(2
\pi)$, in Hz. Squares : signal (first heterodyne term). Circles :
ghost (second heterodyne term).

  \item Squared amplitude of the instrumental response defined by
eq. \ref{eq_B_omega_defn}. Camera framerate : $\omegaS / 2 \pi = 8$
Hz. Exposure time : $\TE = 124 \,\rm ms$. $n=4$. Representation of
$10 \log_{10} [\left|B_{\pm}(\omegaLO-\omegaL)\right|^2]$, in dB.
Horizontal axis : detuning frequency $(\omegaLO - \omegaL)/(2 \pi)$,
in Hz. Dotted line : $B_+$. Continuous line : $B_-$.

  \item Frequency diagram of heterodyne terms spectral components
in the case where the detuning frequency is set to $\omegaLO -
\omegaL =  \Delta \omega - \omegaS/n$.

  \item Images ($1024 \times 1024$ pixels) of the sample for
$(\omegaLO - \omegaL)/(2 \pi)$ equal to 0 Hz (a), 400 Hz (b), 4000
Hz (c) and 8000 Hz (80 ms image exposure time,  4-image
demodulation. Arbitrary logarithmic scale display.

  \item Traces obtained by summation along columns of
Fig.\ref{fig_images_80ms_4im}(a) to (d) intensities. Curves a to d
correspond to a detuning frequency $(\omegaLO - \omegaL) / (2 \pi)$
equal to 0 Hz (a), 400 Hz (b), 4000 Hz (c) and 8000 Hz (d).
Horizontal scale is the image horizontal pixel index. Vertical scale
is in linear arbitrary units.

  \item Frequency spectra of the light diffused through a
suspension of latex particles in brownian motion. Exposure time is
$\TE=80$ ms. Demodulation is performed with 4 (a), 8 (b), 16 (c) and
32 (d) images. Horizontal axis is the detuning frequency $(\omegaLO
- \omegaL)/(2\pi)$ in kHz. Vertical scale is in linear arbitrary
units. The four curves overlap.

  \item Spectra measured with exposure time $\TE=80$ ms (a), 20 ms
(b), 5 ms (c), and 4-image demodulation. Note that curve (a) the
same as Fig.\ref{fig_spectrum_80ms} (4 images and 80 ms). Vertical
scale is in linear arbitrary units. The three curves overlap.

  \item Frequency lineshapes of the light diffused through the cell
for different concentrations of latex spheres. Exposure time is $\TE
= 80$ ms. 4-image demodulation. Horizontal axis is the detuning
frequency $(\omegaLO - \omegaL)/(2\pi)$ in kHz. Vertical scale is in
linear arbitrary units. Volumic concentration of latex beads : $2.9
\times 10^{-3}$ (a), $1.5 \times 10^{-3}$ (b), $7.3 \times 10^{-4}$
(c), to $3.6 \times 10^{-4}$ (d).

\end{enumerate}

\end{document}